\documentclass[journal]{IEEEtran}

\usepackage{cite}

\ifCLASSINFOpdf
\else
\fi

\usepackage{amsmath}
\usepackage{algorithmic}
\usepackage{array}
\usepackage{stfloats}
\usepackage{xcolor}
\usepackage{indentfirst}
\usepackage{epsfig,setspace}
\usepackage{float,algorithm}
\usepackage{bm}
\usepackage{multirow, hhline}
\usepackage{makecell}
\usepackage{amssymb}
\usepackage{pifont}
\usepackage{arydshln}
\usepackage{graphicx}
\usepackage{booktabs}
\newcommand{\cmark}{\ding{51}}
\newcommand{\xmark}{\ding{55}}

\begin{document}

\title{Recent Progress in the CUHK\\Dysarthric Speech Recognition System}


\author{Shansong Liu$^*$, Mengzhe Geng$^*$, Shoukang Hu$^*$, Xurong Xie$^*$, Mingyu Cui, Jianwei Yu, \\Xunying Liu,~\IEEEmembership{Member,~IEEE}, Helen Meng,~\IEEEmembership{Fellow,~IEEE}
    \thanks{$^*$These authors contributed equally.\\
    \indent Shansong Liu, Mengzhe Geng, Shoukang Hu, Mingyu Cui, Jianwei Yu are with the Chinese University of Hong Kong, Hong Kong (email: \{ssliu,mzgeng,skhu,mycui,jwyu\}@se.cuhk.edu.hk).\\
    \indent Xurong Xie is with Chinese University of Hong Kong, Hong Kong, China, and Shenzhen Institutes of Advanced Technology, Chinese Academy
of Sciences, Shenzhen, China (email: xr.xie@siat.ac.cn).\\
    \indent Xunying Liu is with the Chinese University of Hong Kong, Hong Kong and the corresponding author (email: xyliu@se.cuhk.edu.hk).\\
    \indent Helen Meng is with the Chinese University of Hong Kong, Hong Kong (email: hmmeng@se.cuhk.edu.hk).}
}

\markboth{JOURNAL OF \LaTeX\ CLASS FILES, VOL. 14, NO. 8, AUGUST 2015}%
{Shell \MakeLowercase{\textit{et al.}}: nothing goes here}

\maketitle

\begin{abstract}
Despite the rapid progress of automatic speech recognition (ASR) technologies in the past few decades, recognition of disordered speech remains a highly challenging task to date. Disordered speech presents a wide spectrum of challenges to current data intensive deep neural networks (DNNs) based ASR technologies that predominantly target normal speech. This paper presents recent research efforts at the Chinese University of Hong Kong (CUHK) to improve the performance of disordered speech recognition systems on the largest publicly available UASpeech dysarthric speech corpus. A set of novel modelling techniques including neural architectural search, data augmentation using spectra-temporal perturbation, model based speaker adaptation and cross-domain generation of visual features within an audio-visual speech recognition (AVSR) system framework were employed to address the above challenges. The combination of these techniques produced the lowest published word error rate (WER) of 25.21\% on the UASpeech test set 16 dysarthric speakers, and an overall WER reduction of 5.4\% absolute (17.6\% relative) over the CUHK 2018 dysarthric speech recognition system featuring a 6-way DNN system combination and cross adaptation of out-of-domain normal speech data trained systems. Bayesian model adaptation further allows rapid adaptation to individual dysarthric speakers to be performed using as little as 3.06 seconds of speech. The efficacy of these techniques were further demonstrated on a CUDYS Cantonese dysarthric speech recognition task.
\end{abstract}


%
\IEEEpeerreviewmaketitle

\vspace{-0.2cm}
\section{Introduction}
\label{sec:intro}

\IEEEPARstart{D}ESPITE the rapid progress of automatic speech recognition (ASR) technologies in the past few decades, recognition of disordered speech remains a highly challenging task to date. Speech disorders such as dysarthria affect millions of people around the world and introduce a negative impact on their quality of life. Speech disorders are caused by a range of neuro-motor conditions including cerebral palsy~\cite{whitehill2000speech}, amyotrophic lateral sclerosis~\cite{makkonen2018speech}, Parkinson disease~\cite{scott1983speech}, stroke or traumatic brain injuries~\cite{jerntorp1992stroke}. Common forms of speech disorders such as dysarthria manifest themselves in neuro-motor problems leading to weakness or paralysis of muscles that are used in articulation~\cite{kent2000research}. This reduces the intelligibility of the resulting speech for human listeners. As the underlying condition deteriorates, people suffering from speech disorders will not only lose their ability to express themselves but also to live independently. Such people often experience co-occurring physical disabilities and other medical conditions at the mean time. Their difficulty in using keyboard, mouse and touch screen based user interfaces makes speech controlled assistive technologies more natural alternatives~\cite{hux2000accuracy,young2010difficulties}, even though speech quality is degraded. To this end, in recent years there has been increasing research interest in developing ASR technologies that are suitable for disordered speech~\cite{christensen2013combining,sehgal2015model,yu2018development,liu2019exploiting,hu2019cuhk}. 

Disordered speech presents a wide spectrum of challenges to current deep neural networks (DNNs) based speech recognition technologies that predominantly target normal speech. First, a large mismatch between disordered and normal speech is often observed. Such difference systematically manifests itself in articulatory imprecision, increased dysfluencies, slower speaking rates and reduced volume and clarity. Furthermore, people suffering from speech impairments tend to use shorter utterances based on isolated words and simple commands, due to the fatigue they encounter when speaking, when communicating with their careers. This limits the long range temporal contexts that current DNN based ASR systems designed for normal speech~\cite{xiong2018microsoft,luscher2019,Park2019} can exploit.

\vspace{-0.2cm}
\begin{table}[htbp]
\caption{Description of publicly available dysarthric speech corpora}
\centering
\label{tab:corpus}
\scalebox{0.8}{
\begin{tabular}{c|c|c|c|c} 
\hline\hline
Corpus  & \#Hours & \#Spk. & Vocab. & Year  \\ 
\hline\hline
Cantonese (CUDYS)~\cite{wong2015development}     & 10       & 16     & -     & 2015  \\
Dutch (EST)~\cite{yilmaz2016dutch}             & 6.3      & 16     & -     & 2016  \\
English (Nemours)~\cite{menendez1996nemours}     & 2.5-3    & 11     & -     & 1996  \\
English (UASpeech)~\cite{kim2008dysarthric}    & 102.7     & 29     & 455   & 2008  \\
English (TORGO)~\cite{rudzicz2012torgo}       & 15       & 15     & 1573  & 2012  \\
\hline\hline
\end{tabular}}
\end{table}

A set of publicly available disordered speech corpora are shown in Table~\ref{tab:corpus}. The Nemours~\cite{menendez1996nemours} corpus contains less than 3 hours of speech from 11 speakers. The similar sized Dutch EST database~\cite{yilmaz2016dutch} contains approximately 6 hours of speech. The English TORGO~\cite{rudzicz2012torgo} and Cantonese CUDYS~\cite{wong2015development} corpora are moderately larger. The former contains 15 hours of speech while the latter contains around 10 hours of speech. By far, the largest available and widely used dysarthric speech database, the English UASpeech~\cite{kim2008dysarthric} corpus, contains 102.7 hours of speech\footnote{Audio recordings collected from multiple microphone channels were used.} recorded from 29 speakers based on single word utterances of digits, computer commands, radio alphabet letters, common and uncommon words, among which 16 are dysarthric speakers while the remaining 13 are healthy control speakers. Compared with more widely available normal speech corpora, such as Switchboard conversational telephone speech~\cite{godfrey1992switchboard} or Librispeech~\cite{panayotov2015librispeech} containing hundreds of to thousands of audio data, all the existing disordered speech corpora are much smaller in size. 

Second, the underlying neuro-motor conditions, often compounded with co-occurring physical disabilities, lead to the difficulty in collecting large quantities of disordered speech required for ASR system development. For data intensive deep learning technologies widely used in current speech recognition systems, large quantities of well-matched, in-domain speech data are essential. Finally, the large variation among speakers with diverse impairment characteristics, severity levels and in different stages of speech disorder progression creates large variation in disordered speech data. This presents a further challenge to the robustness of disordered speech recognition systems. For the above reasons, state-of-the-art ASR systems designed for normal speech often produce very high recognition error rate above 50\% when being applied to impaired speech~\cite{hermann2020dysarthric,wang2021improved}.

In order to address these issues, the main part of this paper presents the recent research efforts made at the Chinese University of Hong Kong to significantly improve the performance of current disordered speech recognition systems on the largest available and widely used 102.7-hour UASpeech corpus. A set of purposefully designed modelling techniques were derived to address the aforementioned challenges. Both the description of individual approaches and how they can be integrated together to obtain the best recognition performance are presented. 

First, motivated by the large mismatch between normal and disordered speech, a systematic investigation of neural network architecture designs targeting dysarthric speech recognition is conducted. These include state-of-the-art ASR system architectures based on either a hybrid DNN-HMM framework, for example, sequence discriminatively trained time delay neural networks (TDNNs) with phonetic states output targets~\cite{peddinti2015time,povey2016purely,waibel1989consonant,povey2018semi}, or end-to-end approaches represented by connectionist temporal classification (CTC)~\cite{graves2006connectionist}, attention based encoder-decoder models using listen, attend and spell (LAS)~\cite{chan2016listen} and the recent Pychain end-to-end TDNN~\cite{Shao2020} systems directly modelling grapheme (letter) sequence outputs. A manually designed DNN architecture tailored for the disordered speech data of UASpeech is then proposed. Automatic neural architecture search (NAS) techniques~\cite{elsken2018neural, liu2018darts,xie2018snas,cai2018proxylessnas, hu2020dsnas,hu2021neural} are further used to refine its structural configurations. 

Second, in order to address the data sparsity problem in disordered speech recognition system development, and inspired by the success of data augmentation techniques widely reported in normal speech recognition tasks~\cite{Park2019,nguyen2020improving,song2020specswap}, data augmentation techniques designed to model the spectral-temporal level deviation of disordered speech from normal speech are used. A combined use of speaker independent perturbation of disordered speech and impaired speaker dependent perturbation using normal speech expands the training data quantity by a factor of 4~\cite{geng2020investigation}.

Third, in order to model the large variability among disordered speakers in both the original and augmented data, model based DNN adaptation methods represented by, for example, learning hidden unit contributions (LHUC)~\cite{swietojanski2014learning} based speaker adaptive training (SAT) were further applied. Bayesian speaker adaptation approaches were also employed to facilitate rapid, instantaneous adaptation to individual speakers’ voice characteristics, using as little as 3.06 seconds of speech per speaker, at the onset of their enrollment to systems.

Lastly, inspired by the bi-modal nature of human speech perception and the success of audio-visual speech recognition (AVSR) technologies when being applied to normal speech~\cite{yu2020audio,Yu2020,yu2020audiojournal}, visual information is further incorporated to improve disordered speech recognition performance. In order to address the data sparsity that arises from the difficulty to record large amounts of high quality audio-visual (AV) data, a cross-domain visual feature generation approach~\cite{liu2020exploiting} was developed. High quality AV parallel data based on normal speech recording of the lip reading sentence (LRS2) dataset~\cite{afouras2018deep} was used to build neural AV inversion systems. These were then used to generate visual features for the UASpeech audio data that do not have video recordings available. Cross-domain AV inversion system adaptation was also performed to minimize the mismatch between the LRS2 and UASpeech audio data.

By incorporating all the above techniques, the best recognition system produced an overall word error rate (WER) of 25.21\% on the 22.6-hour UASpeech test set containing 16 dysarthric speakers. To the best of our knowledge, this is the lowest WER published so far on the same task reported in the literature~\cite{christensen2013combining,sehgal2015model,yu2018development,xiong2019phonetic,liu2020exploiting,geng2020investigation}. An overall WER reduction of 5.39\% absolute (17.61\% relative) was obtained over the CUHK 2018 system featuring a 6-way DNN system combination~\cite{yu2018development} which defined state-of-the-art performance at the time. A further set of experiments and performance analysis were then conducted on the Cantonese CUDYS~\cite{wong2015development} corpus which is based on a short sentence recognition task. 

The main contributions of this paper are summarized below:

1) To the best of our knowledge, this is the first work to systematically investigate deep neural network architecture design for disordered speech recognition. In contrast, previous research in this area largely focused on using one single type of expert DNN architecture targeting normal speech~\cite{takashima2019end,hermann2020dysarthric}. Detailed comparison and performance analysis between traditional hybrid DNN-HMM and more recent end-to-end approaches were not conducted in the prior works. In addition, novel auto-configured neural architecture search approaches are proposed in this paper for disordered speech recognition.

2) This paper presents the first work that investigates different data augmentation techniques for disordered speech recognition. Both normal and disordered speech were exploited in the augmentation process and evaluated over a wide range of expert hybrid or end-to-end, manually or automatically designed DNN system architectures. In contrast, previous research focused on using temporal perturbation performed only on normal speech data during augmentation~\cite{vachhani2018data,xiong2019phonetic}.

3) This paper presents the first work on rapid speaker adaptation for disordered speech recognition. The proposed Bayesian DNN adaptation approaches can capture the diverse characteristics among dysarthric speakers using as little as 3.06 seconds of speech. In contrast, previous research focused on batch mode adaptation required significant amounts of speaker level data, for example, over one hour on the UASpeech task~\cite{sharma2013acoustic}.

4) This paper presents the first attempt of using cross-domain visual feature generation for audio-visual disordered speech recognition within a state-of-the-art AVSR system. This is contrast to previous AVSR research on disordered speech where the AV data sparsity was largely unaddressed~\cite{salama2014audio,liu2019exploiting}.

The rest of this paper is organized as follows. The details of system development on the UASpeech task are presented from Sec.~\ref{sec:asr_design} to~\ref{sec:avsr}. Among these, a range of hybrid and end-to-end classic ASR system DNN architectures, together with manually designed DNN and neural architecture search (NAS) auto-configured DNN systems as well as their performance across varying speech disorder severity levels are first shown in Sec.~\ref{sec:asr_design}. Disordered speech data augmentation techniques are then presented in Sec.~\ref{sec:data_aug}. Performance of model based dysarthric speaker adaptation methods are shown in Sec.~\ref{sec:spkr_adapt}. Audio-visual disordered speech recognition systems and their performance are presented in Sec.~\ref{sec:avsr}. Further performance analysis against recent published state-of-the-art systems constructed on the same UASpeech task is conducted in the same section (Sec.~\ref{sec:avsr}). A comparable and smaller set of experiments and performance analysis were then conducted on the Cantonese CUDYS corpus to further confirm the trends previously found on the English UASpeech data. The last section draws the conclusions and discusses possible future works. For all results presented in this paper, matched pairs sentence-segment word error (MAPSSWE) based statistical significance test was performed at a significance level $\alpha=0.05$.
\vspace{-0.2cm}

\section{ASR System Architecture}
\label{sec:asr_design}


In this section, a large set of expert designed hybrid and end-to-end system architectures, together with manually designed DNN and neural architecture search (NAS) auto-configured DNN systems considered in this paper, are extensively evaluated in the experiments of this section on the UASpeech task.


The UASpeech corpus is the largest publicly available disordered speech corpus that is designed as an isolated word recognition task~\cite{kim2008dysarthric}. Approximately 103 hours of speech was recorded from 29 speakers among which 16 are dysarthric speakers while the remaining 13 are healthy control speakers. For speech recognition system development, the entire corpus is further divided into 3 subset blocks per speaker, with each block containing different speech contents based on a mix of common and uncommon words. Among these, the same set of common words contents are used in all three blocks, while the uncommon words in each block are different. The data from Block 1 (B1) and Block 3 (B3) of all the 29 speakers are used as the training set (69.1 hours of audio, 99195 utterances in total), while the data of Block 2 (B2) collected from all the 16 dysarthric speakers (excluding speech from healthy control speakers) serves as the evaluation data set (22.6 hours of audio, 26520 utterances in total). After removing excessive silence at the start and end of speech audio segments~\cite{yu2018development}, a combined total of 30.6 hours of audio data from Block 1 and 3 (99195 utterances) were used as the training set, while 9 hours of speech from Block 2 (26520 utterances) was used for performance evaluation. Following the configurations specified in~\cite{yu2018development,christensen2013learning}, recognition was performed using a uniform language model with a word grammar network.

\begin{table*}
\vspace{-0.3cm}
\centering
\caption{1-best and oracle performance and system description of expert designed neural network architectures (Sys. 1-11), manually designed DNN (Sys. 12) and automatically searched neural architecture (Sys. 13). All systems were trained using 80-dimension input features based on Mel-scale filter banks (FBKs) and delta features. ``Seen" and ``Unseen" denote test set words occurring in the training data or otherwise. ``Very low'', ``Low'', ``Mild'' and ``High'' denote different intelligibility groups.}
\vspace{-0.2cm}
\label{tab:expert_system}
\scalebox{0.7}{\begin{tabular}{c|c|c|c|c|c|c|c|c|c|c|c|c|c} 
\hline\hline
\multirow{2}{*}{Sys.} & \multirow{2}{*}{Model}   & \multirow{2}{*}{Structure}                                                  & \multirow{2}{*}{Obj.}   & \multirow{2}{*}{Tgt.} & \multirow{2}{*}{\#Param} & \multicolumn{7}{c|}{WER\%}                                  & \multirow{2}{*}{\begin{tabular}[c]{@{}c@{}}Oracle\\WER\%\end{tabular}}  \\ 
\cline{7-13}
                      &                          &                                                                             &                         &                       &                          & Seen  & Unseen & Very low & Low   & Mild  & High  & Average &                                                                         \\ 
\hline
1                     & DNN                      & 6-feedforward                                                               & \multirow{4}{*}{CE}     & phn.                  & 25.50M                    & 21.81 & 54.24  & 67.20    & 36.34 & 28.63 & 15.65 & 35.20   & 17.35                                                                   \\ 
\cline{1-3}\cline{5-14}
2                     & \multirow{2}{*}{TDNN}    & \multirow{2}{*}{6-tdnn}                                                     &                         & phn.                  & 25.34M      & 24.17      & 53.99      & 69.38        & 40.28     & 30.29     & 14.68     & 35.80       & 15.42                                                                       \\
3                     &                          &                                                                             &                         & gph.                  & 25.36M                         & 25.95     & 62.75      & 73.84        & 44.13     & 35.17     & 19.47     & 39.98       & 21.32                                                                       \\ 
\cline{1-3}\cline{5-14}
4                     & BLSTM                    & 4-blstm                                                                     &                         & phn.                  & 29.52M                    & 21.41 & 65.62  & 67.31    & 40.96 & 32.69 & 21.41 & 38.50   & 26.87                                                                   \\ 
\hline\hline
5                     & \multirow{2}{*}{CTC}     & \multirow{2}{*}{\begin{tabular}[c]{@{}c@{}}4-conv2d+\\3-blstm\end{tabular}} & \multirow{2}{*}{CTC}    & phn.                  & 25.99M   & 33.43 & 80.06  & 81.44    & 56.65 & 47.72 & 31.51 & 51.72   & 44.03                                                                       \\
6                     &                          &                                                                             &                         & gph.                  &  25.98M                        & 39.37     & 86.80      & 87.05         & 62.42      & 55.33     & 37.78     & 57.98       & 54.05                                                                       \\ 
\hline
7                     & LAS                      & \begin{tabular}[c]{@{}c@{}}4-conv2d+\\3-blstm encoder+\\1-lstm decoder\end{tabular}                                                                           & \multirow{6}{*}{LAS}                     & \multirow{6}{*}{gph.}                  & 27.28M                    & 27.39 & 99.24  & 74.74    & 56.18 & 52.55 & 43.92 & 55.28   & 39.48                                                                       \\ 
\cline{1-3}\cline{6-14}
8                     & \begin{tabular}[c]{@{}c@{}}LAS~\cite{wang2021improved}\\(Librispeech\\ 1000hr) \end{tabular}                        & \begin{tabular}[c]{@{}c@{}}\\6-conv2d+\\5-blstm encoder+\\1-lstm decoder\end{tabular}                                                                           &                      &                   & \multirow{2}{*}{42.78M}                    & 73.43 & 81.00  & 98.80    & 90.70 & 82.90 & 47.20 & 76.40   & -                                                                       \\ 
\cline{1-1}\cline{7-14}
9                     & \begin{tabular}[c]{@{}c@{}}+domain/speaker adapt\end{tabular}                     &                                                                            &                      &                   &                     & 23.03 & 53.55  & 68.70    & 39.00 & 32.50 & 12.20 & 35.00   & -                                                                       \\ 
\hline
10                    & \multirow{2}{*}{Pychain TDNN} & \multirow{2}{*}{6-conv1d}                                                   & \multirow{2}{*}{LF-MMI} & phn.                  & 25.01M   & 26.56     & 39.32      & 66.50        & 36.20     & 24.13     & 10.27     & 31.57       & 12.14                                                                       \\
11                    &                          &                                                                             &                         & gph.                  & 25.84M                         & 27.16     & 62.35      & 72.47        & 46.20     & 34.25     & 20.95     & 40.97       & 17.23                                                                       \\ 
\hline\hline
12                    & Manual DNN               & \multirow{2}{*}{7-feedforward}                                                                           & \multirow{2}{*}{CE}     & \multirow{2}{*}{phn.} & 5.86M                     & 21.94 & 46.18  & 69.82    & 32.61 & 24.53 & 10.40 & \textbf{31.45}   & 15.35                                                                   \\
\cline{1-2}\cline{6-14}
13                    & NAS DNN                 &                                                                            &                         &                       & 4.73M                     & 22.41 & 43.85  & 68.69    & 33.08 & 22.86 & 9.82  & \textbf{30.83}   & 12.08                                                                   \\
\hline\hline
\end{tabular}}
\vspace{-0.3cm}
\end{table*}


The performance of various expert neural architectures based recognition systems are shown in line 1 to 11 of Table~\ref{tab:expert_system} together with the modelling units, structural configurations, model complexity and error cost functions used in training. These systems include the hybrid frame level cross-entropy (CE) trained DNN model~\cite{yu2018development}, with tied tri-phone state targets (Sys. 1), TDNN system~\cite{peddinti2015time,povey2016purely,waibel1989consonant,povey2018semi} with tied tri-phone or tri-grapheme state targets (Sys. 2, 3),  bi-directional long short-term memory (BLSTM) RNN modelling tied tri-phone state targets~\cite{yu2018development} (Sys. 4), and a set of end-to-end systems directly modelling phoneme or grapheme (letter) sequence outputs based on either the CTC~\cite{graves2006connectionist} (Sys. 5, 6), LAS~\cite{chan2016listen,wang2021improved} (Sys. 7, 8, 9), or the Pychain TDNN architecture with untied bi-phone or bi-grapheme outputs~\cite{Shao2020} (Sys. 10, 11). As the UASpeech training data set does not cover all the test data words, direct acoustic to word end-to-end approaches represented by RNN word transducers~\cite{alex2012sequence} are impractical for this task. Hence, the scope of the investigation over possible neural network architectures is restricted to those modelling either sub-word phonetic targets or grapheme labels.


The performance of the baseline CE trained hybrid DNN system modelling tied tri-phone state targets is shown in line 12 of Table~\ref{tab:expert_system}. This baseline system architecture was manually designed by applying a series of modifications on top of the first phonetic hybrid DNN system (Sys. 1 in Table~\ref{tab:expert_system}, also served as one of the component branches in our 2018 UASpeech system using system combination~\cite{yu2018development}). This seed phonetic hybrid DNN (Sys. 1 in Table~\ref{tab:expert_system}), serving as the starting point of our baseline DNN system development, contains 6 hidden layers, each with 2000 neurons using Sigmoid activation functions before the output layer. Acoustic features fed into the network are 80-dimension Mel-scale filter banks (FBKs) and delta features using a context of 9 consecutive frames. Decision tree tied tri-phone states are used and modeled using Softmax function at the output layer. 

\vspace{-0.1cm}
\begin{figure}[htbp]
\centering
    \centerline{\includegraphics[width=8cm,height=4cm]{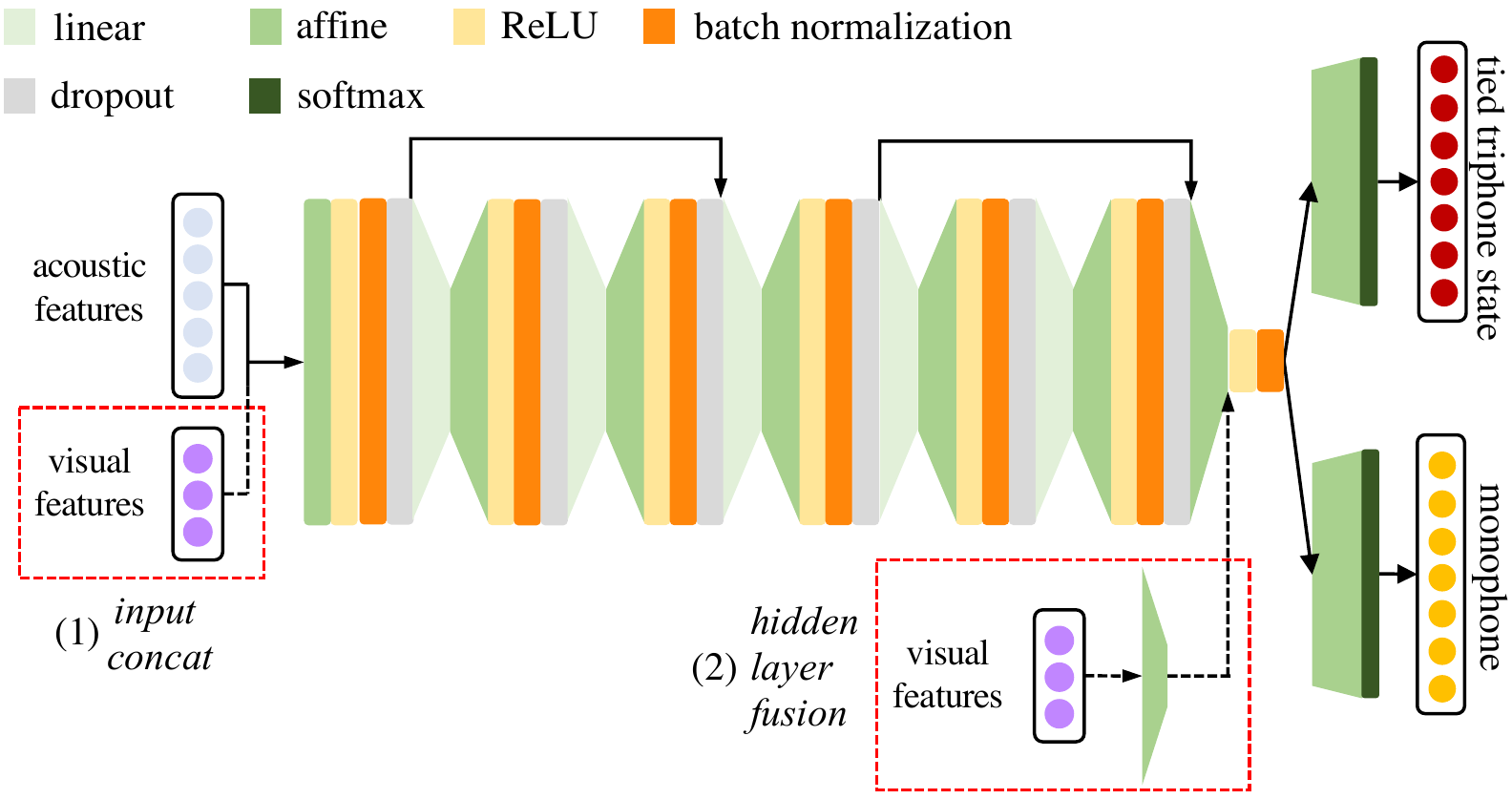}}
    \vspace{-0.3cm}
    \caption{Baseline system architecture of the manually designed DNN. The parts circled by red dotted boxes will be used later for audio-visual speech recognition system construction in Sec.~\ref{sec:avsr}.}
    \label{fig:base_av}
\end{figure}
\vspace{-0.1cm}

A set of architecture modifications are then performed on this 6-layer prototype DNN: 1) a 100-dimension bottleneck hidden layer is inserted immediately before the output layer followed with Sigmoid activation functions, in order to constrain the dimensionality while maintaining the necessary information for tri-phone state classification; 2) in order to address the issues of overfitting and vanishing gradient, a group of neural operations composed of ReLU activation, batch normalization~\cite{ioffe2015batch} and dropout~\cite{srivastava2014dropout} are applied to each hidden layer except the newly added seventh bottleneck hidden layer; 3) in order to reduce the overall number of network parameters given the limited dysarthric speech available in the UASpeech corpus, the weight matrices of the first 5 hidden layers positioned before the ReLU activations were further decomposed into 5 pairs of factored 2000$\times$200 linear and 200$\times$2000 affine matrices, similar with those used in the factored TDNN system~\cite{peddinti2015time,povey2016purely,waibel1989consonant,povey2018semi};  4) in order to compensate the loss of hidden layer information due to the use of subspace factored weight matrices, skipping connections from the first to third, and from the fourth to sixth hidden layers are also added; 5) in order to reduce the risk of overfitting to unreliable tri-phone state alignments during system training, mono-phone alignments are introduced to form a second auxiliary task, with the multi-task weight set as 0.5. After the aforementioned modifications, the resulting manually designed baseline DNN architecture is shown in Fig.~\ref{fig:base_av} (minus the fusion with video information in dotted parts later used for AVSR systems).



Several trends can be observed from Table~\ref{tab:expert_system}. 


1) The hybrid LSTM system (Sys. 4), CTC and LAS based end-to-end systems (Sys. 5, 6, 7) traditionally designed for learning longer temporal contexts are outperformed by the comparable hybrid DNN, TDNN and Pychain TDNN systems (Sys. 1, 2, 3, 10, 11) modelling more restricted contexts.

2) As the training data does not cover all the test data words, the CTC and LAS systems (Sys. 5, 6, 7) produced a large disparity on WER between the seen and unseen words than the other systems in Table~\ref{tab:expert_system}. This may be in part attributed to the poor generalization of CTC and LAS systems when constructed using only the words found in UASpeech training data. Similar performance rank in terms of oracle error rates (last column in Table~\ref{tab:expert_system}) between the CTC and LAS systems against other systems in the table can also be found.

3) The above observations may be attributed to a combination of two factors: the comparatively more limited training data size and the mismatch against normal speech that manifests in, for example, the shorter utterance duration of approximately 3 seconds on average in the UASpeech data. In order to assess the impact from data quantity on the end-to-end systems performance, further experiments are then conducted. A 1000-hour Librispeech~\cite{panayotov2015librispeech} (normal speech) data trained larger size (42.78M parameters) LAS system (Sys. 8) gives a WER of 76.4\%~\cite{wang2021improved} when directly used to recognize the UASpeech data. After domain and speaker adaptation, this cross-domain adapted LAS system’s WER (Sys. 9) was reduced to 35.0\%, on par with the UASpeech data trained hybrid feedforward DNN system (Sys. 1) while still significantly ($\alpha=0.05$) outperformed by the Pychain TDNN system (Sys. 10) using no out-of-domain speech data by 3.43\% absolute in WER.


4) Compared with the starting point DNN system (Sys. 1) in table~\ref{tab:expert_system}, this baseline manually designed hybrid DNN (Sys. 12) produced an overall absolute WER reduction of 3.75\%, as well as 77\% relative reduction in model size. A similar model size reduction ratio was also obtained over various other hybrid and end-to-end systems (Sys. 2 to 7, 10, 11) in Table~\ref{tab:expert_system}. Based on its performance and model compactness, the baseline DNN system architecture (Sys. 12) in Table~\ref{tab:expert_system} was used in the following neural architecture search experiments in the rest of this section to automatically learn the optimal subspace projection dimensionality at each hidden layer.

Neural architecture search (NAS) techniques~\cite{elsken2018neural} can efficiently automate neural network structure designs that have been largely based on expert knowledge or empirical choice to date. Among existing NAS methods, differentiable neural architecture search (DARTS)~\cite{liu2018darts,xie2018snas,cai2018proxylessnas,hu2020dsnas,hu2021neural} benefits from a distinct advantage of being able to simultaneously compare a very large number of candidate architectures during search time. This is contrast to earlier and more expensive forms of NAS techniques based on, for example, genetic algorithms~\cite{stanley2002evolving} and Reinforcement learning (RL)~\cite{zoph2016neural, pham2018efficient}, where explicit system training and evaluation are required for a large number of candidate structures under consideration. 

Architecture search using DARTS is performed over an over-parameterized parent super-network containing paths connecting all candidate DNN structures to be considered. The search is transformed into the estimation of the weights assigned to each candidate neural architecture within the super-network. The optimal architecture is obtained by pruning lower weighted paths. This allows both architecture selection and candidate DNN parameters to be consistently optimized within the same super-network model. 




With no loss of generality, we introduce the general form of DARTS architecture selection methods. For example, the $l$-th layer output $\mathbf{h}^l$ can be computed as follows in the super-network: 
\begin{equation}
\label{eq:darts_fw}
\setlength{\abovedisplayskip}{4pt}
\setlength{\belowdisplayskip}{4pt}
\begin{aligned}
\mathbf{h}^l=&\sum_{i=0}^{N^l-1}\!\lambda_i^l\phi_i^l(\mathbf{W}_i^{l}\mathbf{z}_{i}^{l-1})
\end{aligned}
\end{equation}

\noindent
where $l$ is the layer index, $\lambda_i^l$, $\mathbf{z}_{i}^{l}$ denote the architecture weight and input vector of the $i$-th candidate choice in layer $l$. $N^l$ is the total number of choices of the layer $l$. The precise forms of neural architectures being considered at this layer is determined by the linear transformation parameter $\mathbf{W}_i^{l}$ and activation function $\phi_i^l(\cdot)$ used by each candidate system. 




In conventional DARTS super-networks, Softmax functions are used to model the architecture selection weight $\lambda_i^l$. When the DARTS super-network containing both architecture weights and normal DNN parameters is trained to convergence, the optimal architecture can be obtained by pruning lower weighted architectures that are considered less important. However, when similar architecture weights are obtained using a flattened Softmax function, the confusion over different candidate systems increases and search errors may occur.

In order to address the above issue, Gumbel-Softmax function~\cite{maddison2016concrete} is used in this paper to sharpen the distribution of architecture weights so that approximately one-hot vectors encoded 1 out of N selection decisions will be obtained. This allows the confusion of choosing different architectures to be minimized. The architecture weights of the Gumbel-Softmax DARTS super-network are computed as, 
\begin{equation}
\label{eq:gumbel_reparam}
\begin{aligned}
\lambda_{i}^{l}=\frac{\exp((\log\alpha_i^l+G_{i}^l)/T)}{\sum_{j=0}^{N^l-1}\exp((\log\alpha_j^l+G_{j}^l)/T)}
\end{aligned}
\end{equation}

\noindent
where $G_{i}^l=-\log(-\log(U_{i}^l))$ is a Gumbel variable, and $U_{i}^l$ is a uniform random variable. When the temperature parameter $T$ approaches 0, the Gumbel-Softmax distribution is close to a categorical distribution~\cite{maddison2016concrete}. The temperature parameter $T$ in the Gumbel-Softmax distribution is annealed from $1$ to $0.03$ throughout our NAS experiments in this paper. 

When using the back-propagation algorithm to update the architecture weight parameters, different samples of the uniform random variable $U_{i}^l$ lead to different values of $\lambda_i^l$ in Eq.~\ref{eq:gumbel_reparam}. The loss function gradient, in a general form for both CE and LF-MMI criteria,  w.r.t $\log\alpha_k^l$ is computed as an average over $J$ samples of the architecture weights, 
\begin{equation}
\label{eq:gumbel_bp}
\setlength{\abovedisplayskip}{4pt} 
\setlength{\belowdisplayskip}{4pt}
\begin{aligned}
\frac{\partial\mathcal{L}}{\partial\log\alpha_k^l}\!=\!\frac{1}{J}\!\sum_{j=0}^{J}\frac{\partial\mathcal L}{\partial\mathbf{h}^{l,j}}\!\sum_{i=0}^{N^l-1}\!\frac{1_{i=k}\lambda_{i}^{l,j}-\lambda_{i}^{l,j}\lambda_{k}^{l,j}}{T}\!\phi_i^l(\mathbf{W}_i^{l}\mathbf{h}^{l-1,j})
\end{aligned}
\end{equation}
\noindent
where $\mathbf{\lambda}^{l,j}$ is the $j$-th sample weights vector drawn from the Gumbel-Softmax distrbution in the $l$-th layer, $\mathbf{h}^{l,j}$ is the output of $l$-th layer by using the $j$-th sample $\mathbf{\lambda}^{l,j}$. The Gumbel-Softmax variables $\mathbf{\lambda}^{l}$ at different layers are assumed to be mutually independent during the sampling. 

In order to find a trade-off between the model performance and complexity, a penalized term is further added to the loss function by incorporating the candidate network sizes, 
\begin{equation}
\label{eq:penalized_darts}
\setlength{\abovedisplayskip}{4pt} 
\setlength{\belowdisplayskip}{4pt}
\begin{aligned}
\mathcal{L} = \mathcal{L}_{MTL} + \eta \sum_{l,i}\lambda_{i}^{l}C_{i}^l,
\end{aligned}
\end{equation}
where $C_{i}^l$ is the number of parameters of the $i$-th candidate considered at the $l$-th layer, and $\eta$ is the penalty scaling factor.

\begin{figure}[t]
    \setlength{\abovedisplayskip}{4pt}
    \setlength{\belowdisplayskip}{4pt}
    \centering
    \includegraphics[width=3.0in]{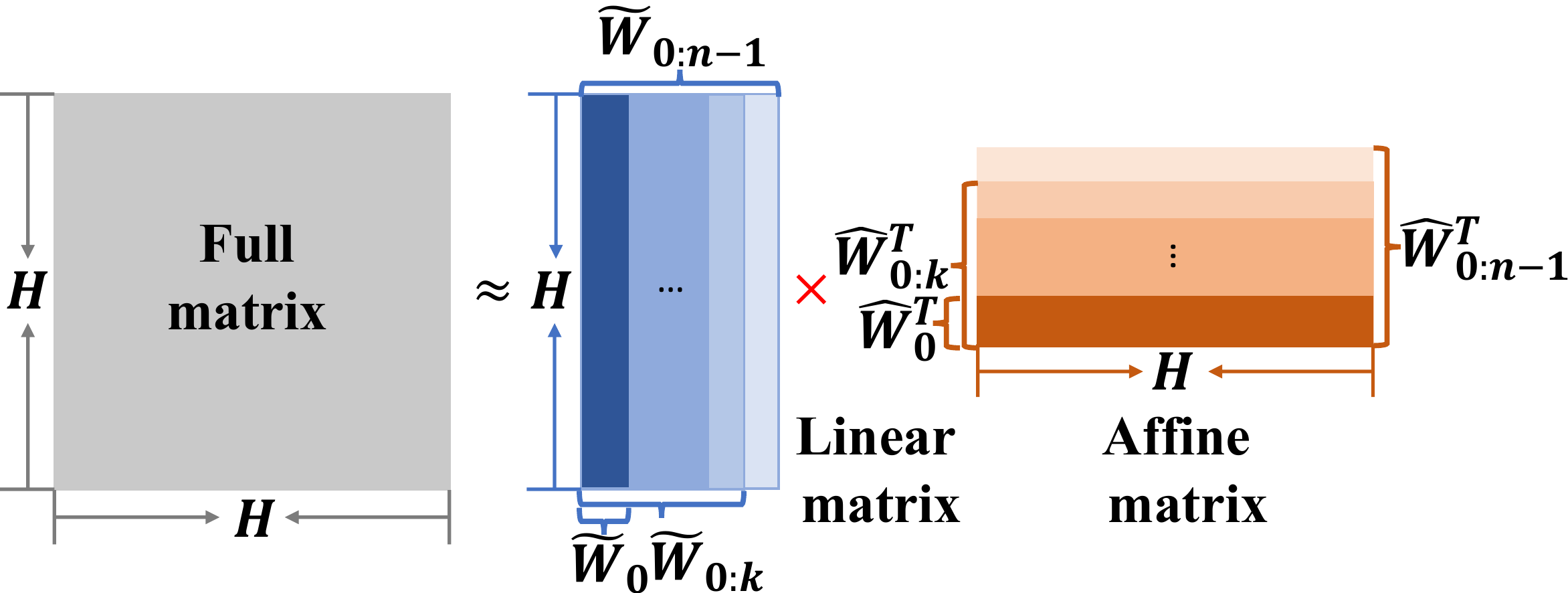}
    \caption{Example part of a super-network containing different bottleneck projection dimensionality choices in one DNN hidden layer.}
    \label{fig:layer_bottleneck_dim_search_space}
    \vspace{-0.3cm}
\end{figure}

In order to facilitate efficient search over a large number of candidate architectures with varying hidden layer specific projection dimensionality settings, parameter sharing among candidate architectures is also used.  An example portion of a DARTS super-network containing all the candidate architectures with different projection dimensions is shown in Fig.~\ref{fig:layer_bottleneck_dim_search_space}. As this portion of super-network is positioned between the decomposed projection and affine linear layers, the activation function $\phi_i^l(\cdot)$ in Eqn. (\ref{eq:darts_fw}) is set as an identity matrix. Parameter sharing among different candidate architectures' linear matrices $\widetilde{\mathbf{W}}_{0:k}$ (left to right from the first column) and affine matrices $\widehat{\mathbf{W}}_{0:k}$ (bottom to up from the first row) ($k \in [0,n-1]$) is implemented by the corresponding submatrices extracted from the large matrix $\widetilde{\mathbf{W}}_{0:n-1}$ and $\widehat{\mathbf{W}}_{0:n-1}$. Such sharing allows a large number of projection dimensionality choices at each of the hidden layers, e.g., selected from 8 values \{25, 50, 80, 100, 120, 160, 200, 240\}, as considered in this paper, to be simultaneously compared for selection during search. This corresponds to a total of $8^{5}=32768$ candidate DNN systems to be selected from. 

The 1-best auto-configured DNN architecture\footnote{NAS selected projection dimensions at each layer: \{160,160,160,120,120\}. $\eta$ in Eqn.~\ref{eq:penalized_darts} is set to be 0.21.} with 4.73 million parameters produced by the above Gumbel-Softmax DARTS approach is shown in the last line (Sys. 13) in Table~\ref{tab:expert_system}. It not only has 20\% fewer parameters than the manually designed baseline DNN system (Sys. 12), but also statistically significantly ($\alpha=0.05$) reduced the WER by 0.6\% absolute WER. The performance of this NAS DNN system, together with the end-to-end systems (Sys. 5 to 7, 10, 11) and the manual DNN baseline (Sys. 12) in Table~\ref{tab:expert_system} will be further evaluated in the following section using data augmentation techniques. 

\vspace{-0.3cm}
\section{Data Augmentation}
\label{sec:data_aug}

Current deep learning-based speech recognition systems are data and resource intensive. In order to reduce the risk of overfitting when constructing such systems using limited training data, data augmentation methods have been explored in the context of normal speech recognition tasks. By expanding the limited training data using, for example, tempo, vocal tract length or speed perturbation~\cite{kanda2013elastic,jaitly2013vocal,ko2015audio}, spectral distortion and masking~\cite{kanda2013elastic,Park2019}, stochastic feature mapping~\cite{cui2015data}, cross domain feature adaptation~\cite{bell2012transcription}, simulation of noisy and reverberated speech to improve environmental robustness~\cite{ko2017study} and end-to-end back translation in end-to-end systems~\cite{hayashi2018back}, the coverage of the augmented training data and the resulting speech recognition systems’ generalization performance can be improved.    

In contrast, so far only limited research on data augmentation targeting disordered speech recognition has been conducted. Motivated by the temporal level differences between disordered speech and normal speech such as slower speaking rates, recent research in this direction has been largely focused on tempo-stretching~\cite{xiong2019phonetic,vachhani2018data} of normal speech recorded from healthy control speakers. The resulting ``disordered like" speech carrying a slower speaking rate is used to augment the limited dysarthric speech training data. Alternative approaches based on cross-domain DNN adaptation~\cite{christensen2013combining},\cite{yu2018development} and voice conversion~\cite{wang2020end} have also been investigated. 

One issue associated with the above existing approaches is that either only applying a temporal level transformation to the normal speech signals while the spectral envelope remains the same, for example, in tempo-stretching~\cite{vachhani2018data}, or a spectral level transformation, for example, using cross-domain feature adaptation~\cite{christensen2013combining}, is applied while the speech tempo remains unaltered. Hence, data augmentation approaches that can exploit the full spectral-temporal differences between normal and disordered speech are preferred, including speaking rate, articulatory imprecision and changes in formant positions and volume. Furthermore, previous researches mainly focused on transforming out-of-domain normal speech to ``disordered like" speech~\cite{jiao2018simulating,celin2020data}, while data augmentation directly using existing disordered speech data has been very rarely studied.

In this section, a systematic investigation over data augmentation techniques based on various spectral-temporal transformations is conducted for disordered speech recognition. The resulting augmented speech data is produced from two sources: a) spectral-temporal modification of normal speech of control speakers to ``disordered like" speech of a target impaired speaker; and b) spectral-temporal perturbation of existing disordered speech. For each of the two sources, three data augmentation techniques were used. These include i) vocal tract length perturbation (VTLP) designed to only alter the spectral envelope to simulate different vocal tract lengths potentially resulted from imprecise articulators’ movements while keeping the speech duration fixed; ii) tempo perturbation modifying the utterance duration to emulate the slower speaking rate in disordered speech while keeping the spectral shape and energy unchanged; and iii) speed perturbation that modifies speech signals in terms of both the duration and shape of the spectral envelope. A summary of these three perturbation methods is presented in Table~\ref{tab:data_aug_tech_diff}.

\begin{table}[t]
\centering
\caption{Comparison of the implementation domain and effects of VTLP, Tempo perturbation and Speed perturbation on modified speech signal. ``\cmark'' indicates that change occurs after perturbation.}
\label{tab:data_aug_tech_diff}
\scalebox{0.85}{\begin{tabular}{c|c|c|c} 
\hline\hline
         & VTLP   & Tempo  & Speed  \\ 
\hline\hline
Implement Domain  & frequency      & time      & time      \\
Signal Duration   & unchanged & \cmark  & \cmark  \\
Spectral Envelope & \cmark  & unchanged & \cmark  \\
\hline\hline
\end{tabular}}
\vspace{-0.3cm}
\end{table}

When performing perturbation of the existing disordered speech training data, a set of global perturbation factors, for example, \{0.9,1.1\} in case of VTLP and speed perturbation, were used. In contrast, when modifying the normal speech of control speakers to simulate that of a target impaired speaker, speaker-level perturbation factors were calculated as the average phonetic duration ratios between their respective speech obtained using phoneme alignment analysis~\cite{xiong2019phonetic}. Force alignment using a GMM-HMM system constructed using the HTK toolkit~\cite{young2006htk} was first performed. The resulting frame-level phoneme alignments were then used to compute the disordered speaker specific perturbation factor as $F_{D_j}=\frac{\overline{t_C}}{t_{D_j}}$. Here $D_j$ denotes the $j$-th dysarthric speaker, $\overline{t_C}$ means the average time duration of all control speakers and $t_{D_j}$ is the time duration of dysarthric speaker $D_j$. 




The data augmentation techniques described in this section were implemented to expand the limited dysarthric speech training data while leaving the test set unchanged. The {\tt HCopy} tool provided by HTK~\cite{young2006htk} was used to apply VTLP based frequency scaling. The {\tt tempo} command based on the WSOLA algorithm~\cite{verhelst1993overlap} and {\tt speed} command provided in Sox~\cite{Sox2020} were used for tempo perturbation and speed perturbation respectively. Following~\cite{ko2015audio}, three sets of global perturbation factors, \{0.9,1.1\}, \{0.9,0.95,1.05,1.1\} and \{0.85,0.9,0.95,1.05,1.1,1.15\}, were applied to obtain augmented data based on disordered speech. Speaker dependent perturbation factors discussed above were applied when modifying normal speech to ``disordered like'' speech. 


80-dimension Mel-scale filter bank and delta features were then extracted from the augmented training data. Pitch parameters extracted using the Kaldi toolkit~\cite{povey2011kaldi} consisting of probability of voicing (POV), normalized pitch, delta-log-pitch and their deltas were also incorporated into the feature front-ends~\cite{liu2019use} for subsequent system development in all the following experiments of this paper. The CTC, LAS, Pychain systems (Sys. 5 to 7, 10, 11) in Table~\ref{tab:expert_system}, together with the manually designed DNN baseline (Sys. 12) and the NAS auto-configured DNN system\footnote{Searched over 6 different choices of projection dimensions \{80,120,160,200,240,300\}.} (Sys. 13), were retrained using various augmented data sets and their performance contrast is shown in Table~\ref{tab:data_aug}.

\begin{table*}
\vspace{-0.3cm}
\centering
\caption{Performance of various hybrid and end-to-end systems trained using different data augmentation approaches. ``CTL'' /``DYS'' stands for normal / disordered speech.``Tempo'' /``Speed'' stands for Tempo / Speed perturbation.``2x'',``4x'' and ``6x'' refer to the amount of augmented training data. the \#hours column shows the total quantity of speech derived from the original training data set or augmented training data after silence stripping applied at utterance boundaries.}
\label{tab:data_aug}
\scalebox{0.75}{\begin{tabular}{c|c|c|c|c|c|c|c|c|c|c|c|c|c|c|c} 
\hline\hline
\multirow{2}{*}{Sys.} & \multirow{2}{*}{Model}                                                 & \multirow{2}{*}{Tgt.}  & \multirow{2}{*}{\#Param} & \multirow{2}{*}{Pitch}  & \multicolumn{3}{c|}{Data Augmentation}                             & \multirow{2}{*}{\#Hours} & \multicolumn{7}{c}{WER\%}                                    \\ 
\cline{6-8}\cline{10-16}
                      &                                                                        &                        &                          &                         & Method                 & CTL                 & DYS                 &                          & Seen  & Unseen & Very low & Low   & Mild  & High  & Average  \\ 
\hline\hline
1                     & \multirow{14}{*}{\begin{tabular}[c]{@{}c@{}}Manual\\DNN\end{tabular}}  & \multirow{15}{*}{phn.} & \multirow{14}{*}{5.86M}  & \xmark                   & \multicolumn{3}{c|}{\multirow{2}{*}{\xmark}}                        & \multirow{2}{*}{30.6}    & 21.94 & 46.18  & 69.82    & 32.61 & 24.53 & 10.40 & 31.45    \\
2                     &                                                                        &                        &                          & \cmark                   & \multicolumn{3}{c|}{}                                              &                          & 21.33 & 42.74  & 67.93    & 30.35 & 22.31 & 9.46  & 29.73    \\ 
\cline{1-1}\cline{5-16}
3                     &                                                                        &                        &                          & \multirow{10}{*}{\xmark} & VTLP                   & \multirow{3}{*}{1x} & \multirow{3}{*}{-}  & 48.0                     & 21.55 & 43.99  & 68.68    & 31.84 & 22.71 & 9.48  & 30.35    \\
4                     &                                                                        &                        &                          &                         & Tempo                  &                     &                     & 52.2                     & 23.44 & 44.68  & 70.71    & 32.78 & 25.12 & 10.32 & 31.77    \\
5                     &                                                                        &                        &                          &                         & Speed                  &                     &                     & 52.2                     & 21.45 & 43.02  & 67.52    & 31.55 & 21.96 & 9.57  & 29.92    \\ 
\cline{1-1}\cline{6-16}
6                     &                                                                        &                        &                          &                         & VTLP                   & \multirow{3}{*}{-}  & \multirow{3}{*}{2x} & 65.5                     & 20.85 & 44.10  & 69.98    & 30.08 & 21.39 & 9.65  & 29.97    \\
7                     &                                                                        &                        &                          &                         & Tempo                  &                     &                     & 65.9                     & 21.78 & 45.02  & 69.32    & 31.75 & 23.94 & 10.07 & 30.90    \\
8                     &                                                                        &                        &                          &                         & Speed                  &                     &                     & 65.9                     & 20.80 & 43.71  & 68.43    & 29.60 & 21.37 & 10.44 & 29.79    \\ 
\cline{1-1}\cline{6-16}
9                     &                                                                        &                        &                          &                         & \multirow{4}{*}{Speed} & \multirow{2}{*}{-}  & 4x                  & 100.9                    & 19.95 & 44.22  & 67.20    & 29.86 & 21.45 & 10.04 & 29.47    \\
10                    &                                                                        &                        &                          &                         &                        &                     & 6x                  & 136.7                    & 20.00 & 44.26  & 67.15    & 30.07 & 21.25 & 10.17 & 29.52    \\ 
\cline{1-1}\cline{7-16}
11                    &                                                                        &                        &                          &                         &                        & 2x                  & 2x                  & 130.1                    & 19.86 & 42.46  & 66.45    & 28.95 & 20.37 & 9.62  & 28.73    \\
12                    &                                                                        &                        &                          &                         &                        & 4x                  & 4x                  & 207.5                    & 19.46 & 42.57  & 66.26    & 28.60 & 19.90 & 9.68  & 28.53    \\ 
\cline{1-1}\cline{5-16}
13                    &                                                                        &                        &                          & \multirow{2}{*}{\cmark}  & \multirow{2}{*}{Speed} & 2x                  & 2x                  & 130.1                    & 20.08 & 42.10  & 66.39    & 29.51 & 20.56 & 9.07  & \textbf{28.72}    \\
14                    &                                                                        &                        &                          &                         &                        & 4x                  & 4x                  & 207.5                    & 19.62 & 42.50  & 66.70    & 28.87 & 20.49 & 9.06  & 28.60    \\ 
\cline{1-2}\cline{4-16}
15                    & \begin{tabular}[c]{@{}c@{}}NAS\\DNN\end{tabular}                     &                        & 6.17M                    & \cmark                   & Speed                  & 2x                  & 2x                  & 130.1                    & 20.06 & 41.46  & 65.63    & 29.04 & 20.66 & 9.08  & \textbf{28.46}    \\ 
\hline\hline
16                    & \multirow{2}{*}{CTC}                                                   & phn.                   & 25.99M                   & \multirow{5}{*}{\cmark}  & \multirow{5}{*}{Speed} & \multirow{5}{*}{2x} & \multirow{5}{*}{2x} & \multirow{5}{*}{130.1}   & 31.02 & 82.34  & 79.46    & 54.26 & 47.86 & 32.84 & 51.15    \\
17                    &                                                                        & gph.                   & 25.98M                   &                         &                        &                     &                     &                          & 35.42 & 88.60  & 84.06    & 60.29 & 52.96 & 37.61 & 56.27    \\
\cline{1-4}\cline{10-16}
18                    & LAS                                                                    & gph                    & 27.28M                   &                         &                        &                     &                     &                     & 24.18 & 99.24 & 72.71    & 53.13 & 50.10 & 43.07 & 40.67    \\
\cline{1-4}\cline{10-16}
19                    & \multirow{2}{*}{\begin{tabular}[c]{@{}c@{}}Pychain\\TDNN\end{tabular}} & phn.                   & 25.01M                   &                         &                        &                     &                     &    & 26.37     & 37.10      & 66.34        & 32.98     & 22.90     & 10.63     & 30.58        \\
20                    &                                                                        & gph                    & 25.84M                   &                         &                        &                     &                     &                          & 31.53     & 68.17      & 78.05        & 51.02     & 40.84     & 24.63     & 45.90        \\
\hline\hline
\end{tabular}}
\vspace{-0.3cm}
\end{table*}


Several trends can be observed from the results of Table~\ref{tab:data_aug}.

1) Among all the three data augmentation methods (VTLP, tempo or speed perturbation), taking the manually designed baseline hybrid phonetic DNN system (Sys. 3 to 8) for example, with similar amounts of augmented training data being used, speed perturbation (Sys. 5, 8) consistently outperformed the other two methods being perturbation was applied to either the healthy control speakers’ data (Sys. 3 to 5) or the original dysarthric speech audio (Sys. 6 to 8). 

2) Further experiments conducted on the same manually designed baseline DNN system suggest applying speed perturbation to both healthy control speaker’s data and dysarthric speech (Sys. 11) outperformed applying it only to either of the two subsets of training data (Sys. 5 and Sys. 8). 

3) Further increasing the amounts of augmented data produced by perturbing both healthy control speaker’s data and dysarthric speech from 130.1 hours (Sys. 11, 13) to 207.5 hours (Sys. 12, 14) only led to marginal WER reductions of 0.1\%-0.2\% absolute, with or without using additional pitch features. Hence, the 130.1-hour augmented data set used by the manual designed DNN (Sys. 13) together with its associated pitch features were used and fixed as the training set for all subsequent UASpeech experiments of this paper.

4) The performance comparison between the manually designed DNN baseline (Sys. 13), the NAS auto-configured DNN\footnote{NAS selected projection dimensions at each layer using the 103.1 hour augmented data set: \{240,200,200,200,240\}. $\eta$ in Eqn.~\ref{eq:penalized_darts} is set to be 0.}, CTC, LAS and Pychain TDNN systems (Sys. 15 to 20) suggests the performance ranking order among all systems constructed using the same 130.1-hour augmented data set in Table~\ref{tab:data_aug} is consistent with that previously found in Table~\ref{tab:expert_system} where no data augmentation is used.

The manually designed DNN and NAS auto-configured DNN systems (Sys. 13, 15), both having the lowest average word error rates and most compact model sizes among all systems in Table~\ref{tab:data_aug}, were then selected to conduct the following speaker adaptation and audio-visual recognition experiments in the rest of this paper. 
\vspace{-0.3cm}

\begin{figure}[htbp]
    \centering
    \includegraphics[width=2.5in]{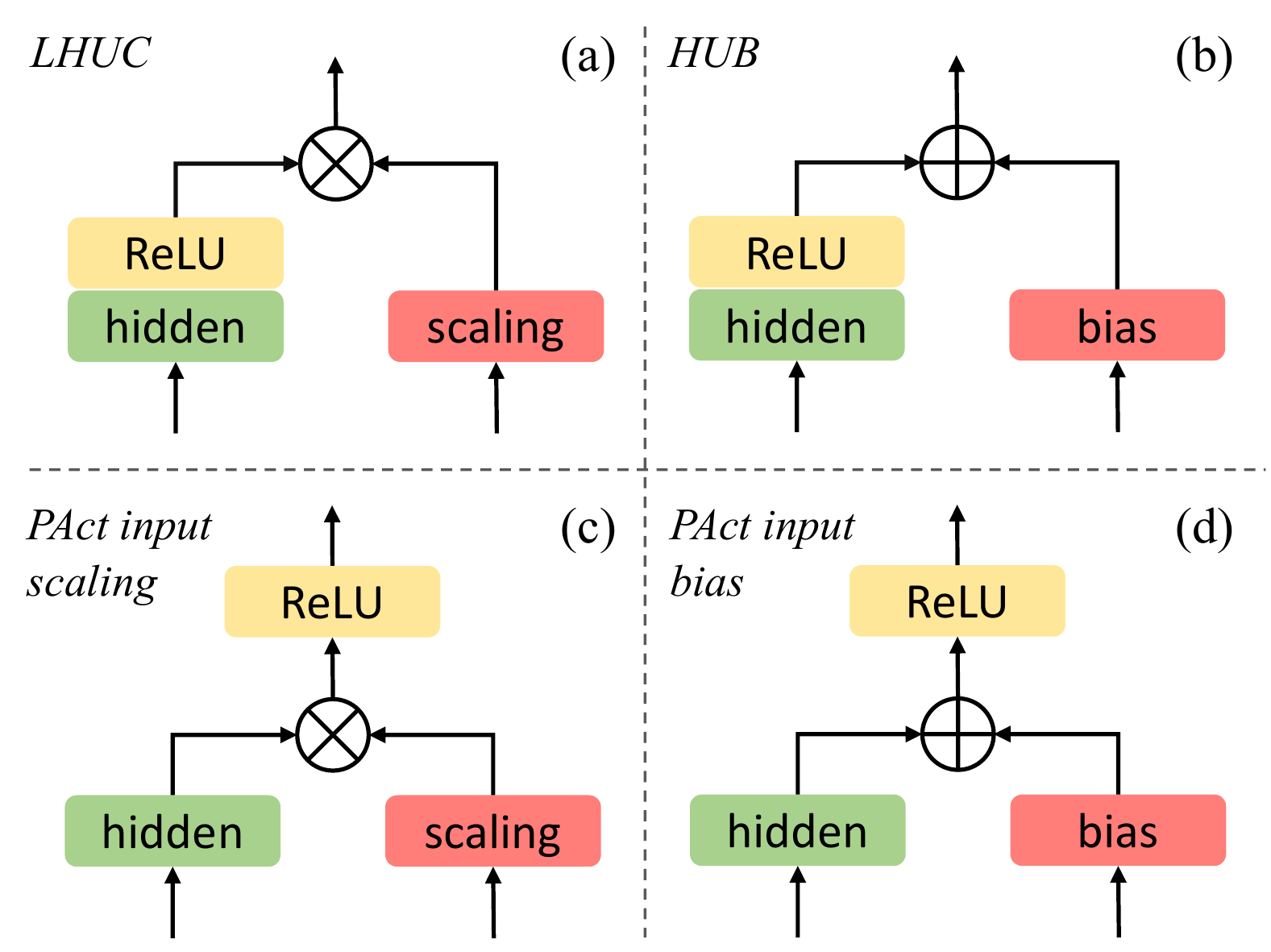}
    \caption{Schematic representation of model based adaptation methods including learning hidden unit contributions (LHUC), hidden unit bias vectors (HUB) and parameterised activation functions (PAct). Speaker dependent scaling or bias parameters are marked in red.}
    \label{fig:LHUC_HUB_PAct}
\end{figure}
\vspace{-0.5cm}

\section{Speaker Adaptation}
\label{sec:spkr_adapt}


A key problem for many speech recognition tasks is to model the systematic and latent variation among diverse speech data. This often creates a large mismatch between the training and evaluation data leading to recognition performance degradation. A major source of such variability is attributable to speaker level characteristics representing factors such as accent and idiosyncrasy, or physiological differences that manifests in, for example, age or gender. For disordered speech recognition tasks considered in this paper, in addition to the wide range of variability factors found in normal speech, the underlying causes and severity levels of speech impairment among dysarthric speakers, also compounded by the spectral-temporal perturbations performed during the data augmentation stage discussed in Sec.~\ref{sec:data_aug}, are expected to further increase the diversity among impaired speakers. 

To this end, speaker adaptation techniques play a vital role in current ASR systems. These can be characterized into several broad categories: a) auxiliary speaker embedding based approaches that encode speaker-dependent (SD) characteristics in a compact vector representation, for example, using i-vectors~\cite{saon2013speaker}; b) feature transformation based methods~\cite{gales1998maximum,uebel1999investigation} that are applied to acoustic front-ends and aim to produce canonical, speaker invariant input features using, for example, feature-space maximum likelihood linear regression (f-MLLR) transforms~\cite{gales1998maximum}; and c) model-based adaptation techniques~\cite{neto1995speaker,swietojanski2014learning,zhang2016dnn,anastasakos1996compact,xie2021bayesian} that exploit specially designed SD DNN model parameters to compensate speaker level variability.  

\vspace{-0.4cm}
\begin{table}[htbp]
\centering
\caption{Performance comparison between different adaptation methods including auxiliary speaker embedding (i-vector) and model based adaptation (LHUC, HUB and PAct). In model based adaptation, the adapted number of hidden layers remain the same during speaker adaptive training and test time adaptation. ``IS’’ and ``IB’’ stand for input scaling and input bias, respectively. ``$\dag$’’ denotes a statistical significance difference is obtained over the system with one adapted hidden layer (Sys. 3 or 15).}
\label{tab:spkr_adapt}
\scalebox{0.75}{\begin{tabular}{c|c|c|c|c|c|c|c|c|c} 
\hline\hline
\multirow{3}{*}{Sys.} & \multirow{3}{*}{Model}                                                 & \multicolumn{2}{c|}{\multirow{3}{*}{\begin{tabular}[c]{@{}c@{}}Adapt\\Method \end{tabular}}} & \multirow{3}{*}{\begin{tabular}[c]{@{}c@{}}Adapt\\Pos. \end{tabular}} & \multicolumn{5}{c}{WER\%}                                                                 \\ 
\cline{6-10}
                      &                                                                        & \multicolumn{2}{c|}{}                                                                        &                                                                       & \begin{tabular}[c]{@{}c@{}}Very \\low \end{tabular} & Low  & Mild & High & Avg.           \\ 
\hline\hline
1                     & \multirow{12}{*}{\begin{tabular}[c]{@{}c@{}}Manual\\DNN \end{tabular}} & \multicolumn{2}{c|}{\xmark}                                                                   & \multirow{2}{*}{\xmark}                                                & 66.4                                                & 29.5 & 20.6 & 9.1  & 28.7           \\ 
\cline{1-1}\cline{3-4}\cline{6-10}
2                     &                                                                        & \multicolumn{2}{c|}{i-vector}                                                                &                                                                       & 66.4                                                & 29.6 & 19.9 & 8.1  & 28.3           \\ 
\cline{1-1}\cline{3-10}
3                     &                                                                        & \multicolumn{2}{c|}{\multirow{7}{*}{LHUC}}                                                   & 1                                                                     & 63.3                                                & 28.2 & 17.7 & 7.7  & 26.7           \\
4                     &                                                                        & \multicolumn{2}{c|}{}                                                                        & 1-2                                                                   & 67.1                                                & 27.7 & 17.8 & 8.1  & 27.5           \\
5                     &                                                                        & \multicolumn{2}{c|}{}                                                                        & 1-3                                                                   & 64.2                                                & 27.9 & 18.0 & 7.6  & 26.8           \\
6                     &                                                                        & \multicolumn{2}{c|}{}                                                                        & 1-4                                                                   & 63.1                                                & 26.5 & 18.4 & 7.8  & 26.4           \\
7                     &                                                                        & \multicolumn{2}{c|}{}                                                                        & 1-5                                                                   & 65.4                                                & 26.8 & 18.0 & 7.9  & 26.9           \\
8                     &                                                                        & \multicolumn{2}{c|}{}                                                                        & 1-6                                                                   & 63.8                                                & 26.8 & 17.0 & 7.8  & 26.4           \\
9                     &                                                                        & \multicolumn{2}{c|}{}                                                                        & 1-7                                                                   & 62.4                                                & 26.8 & 16.5 & 7.5  & \textbf{25.9}$^\dag$  \\ 
\cline{1-1}\cline{3-10}
10                    &                                                                        & \multicolumn{2}{c|}{HUB}                                                                     & \multirow{3}{*}{1-7}                                                  & 63.6                                                & 27.8 & 19.1 & 8.6  & 27.2           \\ 
\cline{1-1}\cline{3-4}\cline{6-10}
11                    &                                                                        & \multirow{2}{*}{PAct} & IS                                                                   &                                                                       & 65.3                                                & 27.2 & 17.8 & 7.6  & 26.9           \\ 
\cline{1-1}\cline{4-4}\cline{6-10}
12                    &                                                                        &                       & IB                                                                   &                                                                       & 64.1                                                & 28.9 & 18.4 & 8.2  & 27.4           \\ 
\hline\hline
13                    & \multirow{12}{*}{\begin{tabular}[c]{@{}c@{}}NAS\\DNN \end{tabular}}  & \multicolumn{2}{c|}{\xmark}                                                                   & \multirow{2}{*}{\xmark}                                                & 65.6                                                & 29.0 & 20.7 & 9.1  & 28.5           \\ 
\cline{1-1}\cline{3-4}\cline{6-10}
14                    &                                                                        & \multicolumn{2}{c|}{i-vector}                                                                &                                                                       & 66.2                                                & 28.8 & 18.3 & 8.1  & 27.7           \\ 
\cline{1-1}\cline{3-10}
15                    &                                                                        & \multicolumn{2}{c|}{\multirow{7}{*}{LHUC}}                                                   & 1                                                                     & 60.7                                                & 27.0 & 18.0 & 8.0  & 26.0           \\
16                    &                                                                        & \multicolumn{2}{c|}{}                                                                        & 1-2                                                                   & 62.5                                                & 26.7 & 17.1 & 7.4  & 25.9           \\
17                    &                                                                        & \multicolumn{2}{c|}{}                                                                        & 1-3                                                                   & 61.4                                                & 26.7 & 16.2 & 7.6  & \textbf{25.6}$^\dag$  \\
18                    &                                                                        & \multicolumn{2}{c|}{}                                                                        & 1-4                                                                   & 61.7                                               & 26.2 & 16.5 & 7.7  & 25.6           \\
19                    &                                                                        & \multicolumn{2}{c|}{}                                                                        & 1-5                                                                   & 61.6                                                & 26.1 & 17.4 & 7.9  & 25.8           \\
20                    &                                                                        & \multicolumn{2}{c|}{}                                                                        & 1-6                                                                   & 62.0                                                & 26.7 & 18.2 & 7.9  & 26.2           \\
21                    &                                                                        & \multicolumn{2}{c|}{}                                                                        & 1-7                                                                   & 63.8                                                & 28.6 & 18.5 & 8.4  & 27.3           \\ 
\cline{1-1}\cline{3-10}
22                    &                                                                        & \multicolumn{2}{c|}{HUB}                                                                     & \multirow{3}{*}{1-3}                                                  & 63.1                                                & 27.7 & 18.9 & 8.6  & 27.1           \\ 
\cline{1-1}\cline{3-4}\cline{6-10}
23                    &                                                                        & \multirow{2}{*}{PAct} & IS                                                                   &                                                                       & 61.5                                                & 27.1 & 16.7 & 7.5  & 25.8           \\ 
\cline{1-1}\cline{4-4}\cline{6-10}
24                    &                                                                        &                       & IB                                                                   &                                                                       & 63.2                                                & 27.5 & 17.6 & 7.6  & 26.4           \\
\hline\hline
\end{tabular}}
\end{table}
\vspace{-0.2cm}

Compared with auxiliary feature and feature transformation based adaptation approaches, model-based adaptation methods have two advantages. First, the SD parameters can be jointly estimated together with the speaker independent parameters in the system consistently in speaker adaptive training (SAT)~\cite{anastasakos1996compact}, in contrast to the often offline, separate estimation of i-vectors and f-MLLR transforms. Second, the amount of speaker specific adaptation data practically determines the SD parameters’ modelling granularity. When a larger amount of speaker level adaptation data is available, the SD parameter’s modelling resolution can be accordingly increased. This allows the trade-off between adapting only parts of the whole acoustic model and building complete speaker dependent systems to be flexibly adjusted. In contrast, for auxiliary feature and feature transform based adaptation, SD feature embedding and transforms of fixed sizes are often used.

In this section, several model-based adaptation approaches are first used to construct speaker adaptively trained disordered speech recognition systems based on the manually designed and auto-configured NAS DNN systems (Sys. 13 and 15 of Table~\ref{tab:data_aug}) in Sec.~\ref{sec:data_aug} using the associated augmented data (using speed perturbation of both control and dysarthric speech). SD transforms that are applied to various parts of the DNN acoustic model, include a) learning hidden unit contributions (LHUC) scaling vectors applied to the hidden layer outputs for each target speaker~\cite{neto1995speaker}; b) parameterized activation functions (PAct) with speaker level vector scaling or bias applied to the input feature before fed into the ReLU  activations~\cite{zhang2016dnn}; and c) hidden unit bias vectors (HUB)~\cite{xie2021bayesian} adding speaker level offset vectors to the hidden unit outputs. The differences between these three model-based adaptation methods when being applied to the hidden layer ReLU activations inside the manually designed and auto-configured NAS DNN systems of line 13 and 15 of Table~\ref{tab:data_aug} in Sec.~\ref{sec:data_aug}, are illustrated in Fig.~\ref{fig:LHUC_HUB_PAct}.

The performance of different speaker adaptation techniques when applied to the manually designed DNN and NAS auto-configured DNN systems are shown in Table~\ref{tab:spkr_adapt}. In all the speaker adaptation experiments of this section, the 1-best outputs produced by the un-adapted speaker independent systems (Sys. 1, 13 in Table~\ref{tab:spkr_adapt}, and earlier in Table~\ref{tab:data_aug} as Sys. 13, 15) served as the supervision for subsequent test time adaptation of speaker adaptively trained (SAT) systems constructed using various adaptation methods introduced in this section. Considering the average amount of speaker specific data used in these experiments is approximately 34 minutes (after silence stripping) for each dysarthric speaker, considerably larger than that found in other ASR tasks such as the Switchboard corpus, the number of DNN hidden layers on which LHUC, HUB or PAct transforms are applied was also fine-tuned to increase modelling resolution of the SD parameters, and to obtain the best adaptation performance for each technique.

Several trends can be observed from Table~\ref{tab:spkr_adapt}.

1) On both the manually designed DNN and NAS auto-configured DNN systems, all of the model based adaptation methods including LHUC, HUB and PAct (Sys. 3 to 12 and Sys. 15 to 24) consistently outperformed i-vector input feature based adaptation (Sys. 2, 14).

2) Among all three model based adaptation methods, the best performance was obtained using the LHUC adapted manually designed DNN and NAS auto-configured DNN systems (Sys. 9, 17), producing overall statistically significant ($\alpha=0.05$) WER reductions of 2.8\% and 2.9\% respectively over the un-adapted baseline systems (Sys. 1, 13).


\begin{figure*}[htbp]
    \vspace{-0.3cm}
    \centering
    \includegraphics[width=6in]{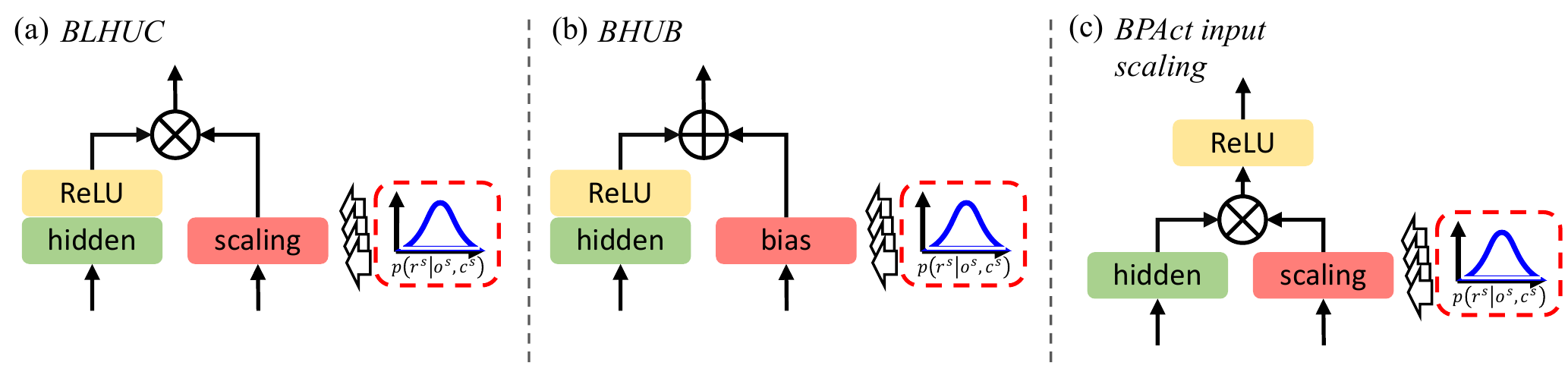}
    \caption{Schematic representation of different Bayesian adaptation methods including Bayesian LHUC (BLHUC), Bayesian HUB (BHUB) and Bayesian PAct (BPAct). The red dotted box at the bottom right corner of each plot represents the SD parameter uncertainty modelled by Bayesian learning.}
    \label{fig:BLHUC_BHUB_BPAct}
    \vspace{-0.3cm}
\end{figure*}

The above speaker adapted systems’ performance were obtained using approximately 34 minutes of adaptation data from each impaired speaker. As discussed in Sec.~\ref{sec:intro}, the underlying neuro-motor conditions, when compounded with co-occurring physical disabilities, lead to the difficulty in collecting large quantities of disordered speech from each target speaker. When developing speech based assistive technologies for such people with speech impairment, it is preferable to employ more powerful adaptation approaches to facilitate rapid, instantaneous adaptation to individual speakers’ voices. However, when performing adaptation using very little speaker level data, for example, a few seconds of speech, a severe data sparsity issue and the resulting modelling uncertainty need to be addressed. 

\begin{table}[htbp]
\vspace{-0.3cm}
\centering
\caption{Performance of baseline or Bayesian adaptation on 130-hour augmented data set trained manually designed DNN and NAS auto-configured DNN systems using varying reducing amounts of adaptation data from 80\% down to only 1 utterance of speech from each target dysarthric speaker (duration in brackets). the bold results indicate the smallest amounts of speaker level adaptation data that can produce performance improvements over the un-adapted systems (sys. 1 or 15) after speaker adaptation. ``$\dag$’’ denotes a statistical significance difference over the un-adapted systems.}
\label{tab:bayesian_adapt}
\scalebox{0.54}{\begin{tabular}{c|c|c|c|c|c|c|c|c|c|c|c} 
\hline\hline
\multirow{3}{*}{Sys.} & \multirow{3}{*}{Model}                                                & \multirow{3}{*}{\begin{tabular}[c]{@{}c@{}}Adapt\\Method\end{tabular}} & \multirow{3}{*}{\begin{tabular}[c]{@{}c@{}}Apt.\\Pos.\end{tabular}} & \multicolumn{8}{c}{WER\% w.r.t. adapt amounts (utt num)}        \\ 
\cline{5-12}
                      &                                                                       &                                                                        &                                                                           & \begin{tabular}[c]{@{}c@{}}1 Utt\\(3.06s)\end{tabular} & \begin{tabular}[c]{@{}c@{}}1\%\\(50.8s)\end{tabular} & \begin{tabular}[c]{@{}c@{}}5\%\\(4.2m)\end{tabular} & \begin{tabular}[c]{@{}c@{}}10\%\\(8.5m)\end{tabular} & \begin{tabular}[c]{@{}c@{}}20\%\\(17.0m)\end{tabular} & \begin{tabular}[c]{@{}c@{}}40\%\\(33.9m)\end{tabular} & \begin{tabular}[c]{@{}c@{}}80\%\\(1.1h)\end{tabular} & \begin{tabular}[c]{@{}c@{}}all\\(1.4h)\end{tabular}  \\ 
\hline\hline
1                     & \multirow{14}{*}{\begin{tabular}[c]{@{}c@{}}Manual\\DNN\end{tabular}} & \xmark                                                                  & \multirow{2}{*}{\xmark}                                                    & \multicolumn{7}{c|}{\multirow{2}{*}{-}}                                                                                                                                                                                                                                                                                                                                                      & 28.7                                                 \\ 
\cline{1-1}\cline{3-3}\cline{12-12}
2                     &                                                                       & i-vector                                                               &                                                                           & \multicolumn{7}{c|}{}                                                                                                                                                                                                                                                                                                                                                                        & 28.3                                                 \\ 
\cline{1-1}\cline{3-12}
3                     &                                                                       & LHUC                                                                   & \multirow{6}{*}{1-7}                                                      & 51.6                                              & 45.3                                                 & 37.8                                                & 32.2                                                 & 30.4                                                  & \textbf{27.7}$^\dag$                                         & 25.9                                                 & 25.9                                                 \\
4                     &                                                                       & BLHUC                                                                  &                                                                           & 44.7                                              & 36.2                                                 & 32.4                                                & 30.1                                                 & 29.6                                                  & \textbf{27.8}$^\dag$                                         & 26.7                                                 & 26.6                                                 \\ 
\cline{1-1}\cline{3-3}\cline{5-12}
5                     &                                                                       & HUB                                                                    &                                                                           & \textbf{28.6}                                     & 28.4                                                 & 28.0                                                & 28.0                                                 & 27.6                                                  & 27.5                                                  & 27.3                                                 & 27.2                                                 \\
6                     &                                                                       & BHUB                                                                   &                                                                           & \textbf{27.5}$^\dag$                                     & 27.6                                                 & 27.5                                                & 27.3                                                 & 27.5                                                  & 27.4                                                  & 27.4                                                 & 27.4                                                 \\ 
\cline{1-1}\cline{3-3}\cline{5-12}
7                     &                                                                       & PAct                                                                   &                                                                           & 47.9                                              & 43.7                                                 & 35.7                                                & 32.1                                                 & 30.4                                                  & \textbf{27.9}$^\dag$                                         & 27.2                                                 & 26.9                                                 \\
8                     &                                                                       & BPAct                                                                  &                                                                           & 36.8                                              & 32.8                                                 & 31.6                                                & 30.1                                                 & \textbf{28.7}                                         & 27.3                                                  & 27.0                                                 & 26.6                                                 \\ 
\cline{1-1}\cline{3-12}
9                     &                                                                       & LHUC                                                                   & \multirow{6}{*}{1-3}                                                      & 46.7                                              & 41.8                                                 & 33.5                                                & 31.2                                                 & 29.7                                                  & \textbf{27.7}$^\dag$                                         & 27.0                                                 & 26.8                                                 \\
10                    &                                                                       & BLHUC                                                                  &                                                                           & 40.8                                              & 31.8                                                 & 28.8                                                & \textbf{27.9}$^\dag$                                        & 27.5                                                  & 26.8                                                  & 26.3                                                 & 26.2                                                 \\ 
\cline{1-1}\cline{3-3}\cline{5-12}
11                    &                                                                       & HUB                                                                    &                                                                           & 29.6                                              & 29.4                                                 & \textbf{28.7}                                       & 27.8                                                 & 27.7                                                  & 27.2                                                  & 27.7                                                 & 27.4                                                 \\
12                    &                                                                       & BHUB                                                                   &                                                                           & 29.1                                              & 28.8                                                 & \textbf{28.3}$^\dag$                                       & 27.7                                                 & 28.1                                                  & 27.8                                                  & 27.9                                                 & 27.9                                                 \\ 
\cline{1-1}\cline{3-3}\cline{5-12}
13                    &                                                                       & PAct                                                                   &                                                                           & 47.1                                              & 41.7                                                 & 33.6                                                & 30.9                                                 & 29.0                                                  & \textbf{27.4}$^\dag$                                         & 26.6                                                 & 26.6                                                 \\
14                    &                                                                       & BPAct                                                                  &                                                                           & 35.8                                              & 31.1                                                 & 30.0                                                & \textbf{28.7}                                        & 27.8                                                  & 26.9                                                  & 26.6                                                 & 26.6                                                 \\ 
\hline\hline
15                    & \multirow{8}{*}{\begin{tabular}[c]{@{}c@{}}NAS\\DNN \end{tabular}}  & \xmark                                                                  & \multirow{2}{*}{\xmark}                                                    & \multicolumn{7}{c|}{\multirow{2}{*}{-}}                                                                                                                                                                                                                                                                                                                                                      & 28.5                                                 \\
\cline{1-1}\cline{3-3}\cline{12-12}
16                    &                                                                       & i-vector                                                               &                                                                           & \multicolumn{7}{c|}{}                                                                                                                                                                                                                                                                                                                                                                        & 27.7                                                 \\ 
\cline{1-1}\cline{3-12}
17                    &                                                                       & LHUC                                                                   & \multirow{6}{*}{1-3}                                                      & 44.1                                              & 34.7                                                 & 29.6                                                & \textbf{28.1}$^\dag$                                        & 27.1                                                  & 26.4                                                  & 25.7                                                 & 25.6                                                 \\
18                    &                                                                       & BLHUC                                                                  &                                                                           & 35.7                                              & 31.5                                                 & \textbf{27.9}$^\dag$                                       & 27.2                                                 & 26.5                                                  & 26.2                                                  & 26.1                                                 & 26.0                                                 \\ 
\cline{1-1}\cline{3-3}\cline{5-12}
19                    &                                                                       & HUB                                                                    &                                                                           & 29.9                                              & 28.9                                                 & \textbf{28.2}$^\dag$                                       & 27.7                                                 & 27.9                                                  & 27.4                                                  & 27.1                                                 & 27.1                                                 \\
20                    &                                                                       & BHUB                                                                   &                                                                           & 29.6                                              & \textbf{26.9}$^\dag$                                        & 26.9                                                & 26.7                                                 & 26.8                                                  & 26.7                                                  & 26.8                                                 & 26.7                                                 \\ 
\cline{1-1}\cline{3-3}\cline{5-12}
21                    &                                                                       & PAct                                                                   &                                                                           & 35.4                                              & 31.2                                                 & 29.2                                                & \textbf{27.9}$^\dag$                                        & 27.5                                                  & 26.2                                                  & 25.8                                                 & 25.8                                                 \\
22                    &                                                                       & BPAct                                                                  &                                                                           & 33.6                                              & 30.9                                                 & \textbf{28.2}$^\dag$                                       & 27.2                                                 & 26.5                                                  & 26.2                                                  & 26.1                                                 & 26.0                                                 \\
\hline\hline
\end{tabular}}
\vspace{-0.3cm}
\end{table}

To this end, the inherent SD parameter uncertainty resulted from limited adaptation data is addressed using Bayesian learning approaches. Rather than learning fixed value estimates of the SD LHUC, HUB or PAct parameters using the standard cross-entropy cost function, the following Bayesian predictive inference (Eqn.~\ref{eq:bayesian}) incorporating the SD adaptation parameter uncertainty is used instead. 

\vspace{-0.5cm}
\begin{equation}
P(\tilde{C}^s|\mathbf{\tilde{O}}^s,\mathbf{O}^s,C^s)=\int{P(\tilde{C}^s|\mathbf{\tilde{O}}^s,\mathbf{r}^s)p(\mathbf{r}^s|\mathbf{O}^s,C^s)d\mathbf{r}^s}
\label{eq:bayesian}
\end{equation}
\vspace{-0.3cm}

\noindent
where $\mathbf{O}^s$, $\mathbf{\tilde{O}}^s$ denote the adaptation data and test data for speaker $s$, $C^s$ stands for the corresponding supervision label, $\mathbf{r}^s$ denotes the SD parameters of speaker $s$, and $\tilde{C}^s$ refers to the output states to be inferred.

The key task of Bayesian adaptation is to learn the underlying SD parameter posterior distribution $p(\mathbf{r}^s|\mathbf{O}^s,C^s)$ used to encode modelling uncertainty. This distribution can be efficiently learned and approximated as a multi-variate Gaussian distribution using a variational inference approach combined with parameter sampling~\cite{xie2019blhuc,xie2021bayesian}. For efficiency, the expectation of SD parameters can be used to approximate the Bayesian integral in Eqn.~\ref{eq:bayesian} for inference during recognition time.

\vspace{-0.4cm}
\begin{equation}
P(\tilde{C}^s|\mathbf{\tilde{O}}^s,\mathbf{O}^s,C^s)\approx{P(\tilde{C}^s|\mathbf{\tilde{O}}^s,\mathbb{E}[\mathbf{r}^s|\mathbf{O}^s,C^s])}
\label{eq:bayesian_approx}
\end{equation}

\noindent
where $\mathbb{E}[\cdot]$ denotes the expectation. Other symbols in Eqn.~\ref{eq:bayesian_approx} are the same with those used in Eqn.~\ref{eq:bayesian}. An example of applying Bayesian SD estimation to various model-based adaptation approaches of Table~\ref{tab:spkr_adapt}, leading to Bayesian LHUC (BLHUC), Bayesian HUB (BHUB) and Bayesian PAct (BPAct) respectively, are shown in Fig.~\ref{fig:BLHUC_BHUB_BPAct}.


A series of Bayesian adaptation experiments were then conducted in order to demonstrate the minimum amounts of dysarthric speaker level data that can produce statistically significant ($\alpha=0.05$) recognition performance improvements over the un-adapted baseline systems (Sys. 1, 15 in Table~\ref{tab:bayesian_adapt}). This can help improve the resulting ASR system’s practical deployment when new dysarthric speakers are freshly enrolled to the system. 

During Bayesian adaptation, the SD parameter prior distribution used was empirically set as $\mathcal{N}(0, 0.001)$ for all adaptation methods. The SD adaptation parameters were estimated using either as fixed values in baseline adaptation, or in a Bayesian fashion, while the SI portion of the parameters inherited from the SAT trained systems were kept fixed. Based on the full data set adaptation results in Table~\ref{tab:spkr_adapt} (Sys. 11, 23 vs. Sys. 12, 24), only input scaling based PAct adaptation is considered here together with LHUC and HUB. Performance contrasts between the baseline and Bayesian adaptation methods using varying reduced amounts of speaker level data randomly sampled from 80\% down to as little as 1 single utterance (3.06 seconds of speech on average) are shown in Table~\ref{tab:bayesian_adapt}. The following trends can be found.

1) Irrespective of which of the three model adaptation methods being used, Bayesian adaptation consistently outperform the comparable baseline adaptation using fixed value estimation across varying reduced amounts of speaker level data from 40\% down to 1 single utterance. For example, on the HUB adapted NAS auto-configured DNN systems (Sys. 19, 20), when only 1\% (50.8 seconds) of speaker level data is used, Bayesian HUB adaptation (Sys. 20) significantly ($\alpha=0.05$) outperformed the comparable HUB adapted baseline (Sys. 19) by 2.0\% and the un-adapted system (Sys. 15) by 1.6\% absolute in WER respectively. 

2) The bold numbers in Table~\ref{tab:bayesian_adapt} indicate the minimum amounts of speaker level data that can produce statistically significant ($\alpha=0.05$) recognition performance improvements over the un-adapted baseline systems (Sys. 1, 15) for each adaptation technique. It is clear that for the manually designed DNN system with Bayesian HUB adaptation (Sys. 6), only a single utterance of approximately 3.06 seconds of speech is required to produce a statistically significant ($\alpha=0.05$) WER reduction of 1.2\% absolute over the un-adapted baseline system (Sys. 1).


\vspace{-0.2cm}
\begin{figure}[htbp]
    \centerline{\includegraphics[width=7.5cm,height=4.9cm]{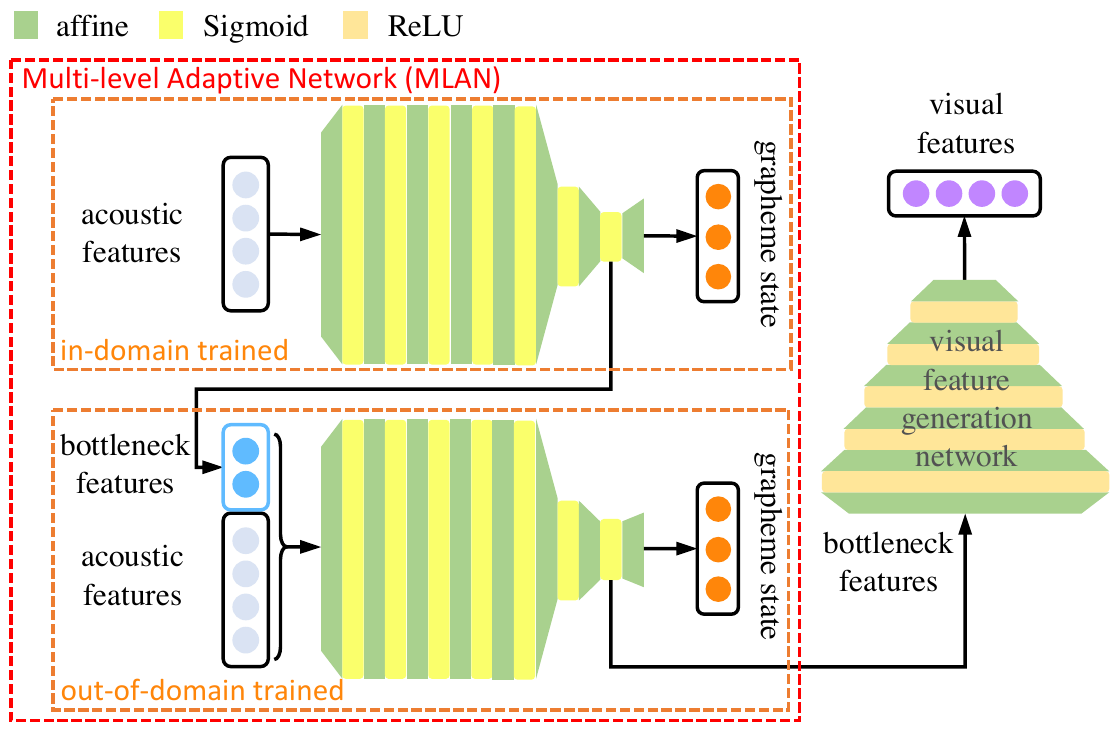}}
    \caption{Cross-domain visual feature generation system. The left part is the MLAN network consisting of two DNN components, while the DNN on the right is the AV inversion model using bottleneck features as cross-domain adapted inputs from the second DNN component of the MLAN network.}
    \label{fig:cross_domain}
\end{figure}
\vspace{-0.5cm}


\section{Audio-visual Speech Recognition}
\label{sec:avsr}

Inspired by the bi-modal nature of human speech perception and the success of audio-visual speech recognition (AVSR) technologies when being applied to normal speech~\cite{yu2020audio,Yu2020,yu2020audiojournal}, visual information is further incorporated to improve disordered speech recognition performance. In order to address the data sparsity resulting from the difficulty to collect large amounts of high quality audio-visual (AV) data, a cross-domain visual feature generation approach~\cite{liu2020exploiting} was developed to generate visual features for the UASpeech original audio data and the augmented audio only data obtained using the speed perturbation method presented in Sec.~\ref{sec:data_aug}. This allows sufficient AV parallel disordered speech data to be used to develop AVSR systems.




High quality AV parallel data based on normal speech recording of the LRS2 dataset~\cite{afouras2018deep} was used to construct AV inversion neural network systems. However, the resulting AV inversion system cannot be directly applied to dysarthric speech given its large mismatch against the normal speech data in the LRS2 corpus. This mismatch may render the generated visual features unreliable to use for subsequent AVSR system development~\cite{su2019cross,liu2020exploiting}. Such mismatch can be compensated using, for example, domain-adversarial neural network (DANN)~\cite{ganin2016domain} or multi-level adaptive network (MLAN)~\cite{bell2012transcription}. Following the comparative analysis over domain adaptation methods for AV inversion in our previous research~\cite{su2019cross}, the MLAN method was adopted to minimize the domain mismatch between the LRS2 and UASpeech audio data.

An example MLAN network consisting of two DNN components is shown in the left portion of Fig.~\ref{fig:cross_domain}. Each component DNN contains a bottleneck layer positioned immediately before the output layer. The MLAN training process includes the following steps: 1) the first-level DNN was trained with the audio data from the in-domain UASpeech corpus; 2) the resulting in-domain dysarthric speech trained DNN was then used to produce bottleneck features for the out-of-domain data of the LRS2 audio; 3) the second-level DNN was trained using the out-of-domain LRS2 audio data concatenated with the bottleneck features computed from the previous step. 

When feedforwarding the UASpeech data into the resulting MLAN network, the combined effect produced by these two cascaded component DNNs is such that the final bottleneck features produced at the second-level DNN component will exhibit smaller mismatch against the bottleneck features obtained by feedforwarding the LSR2 audio into the MLAN network. These cross-domain adapted bottleneck features are used in AV inversion model training and visual feature generation (in the right part of Fig.~\ref{fig:cross_domain}). The dimensionality of these MLAN bottleneck features was set to 80, in line with the settings used in our earlier research~\cite{liu2019exploiting}. The resulting cross-domain adapted AV inversion system was then applied to 103.1-hour augmented training data previously derived using speed perturbation in Sec.~\ref{sec:data_aug} (used by Sys. 13, 15 in Table~\ref{tab:data_aug}), as well as the test set, to produce 25-dimension visual features for AVSR systems development.

\begin{table}[t]
\centering
\caption{Performance of 103.1-hour augmented training data based LHUC SAT adapted audio-only and audio-visual systems using various AV modality fusion: a) input feature concatenation; b) hidden layer fusion; and c) score fusion.}
\label{tab:audio_visual}
\scalebox{0.74}{\begin{tabular}{c|c|c|c|c|c|c|c|c} 
\hline\hline
\multirow{3}{*}{Sys.} & \multirow{3}{*}{Model}                                                & \multirow{3}{*}{Vis.}  & \multirow{3}{*}{AV Fusion} & \multicolumn{5}{c}{WER\%}                      \\ 
\cline{5-9}
                      &                                                                       &                        &                               & \begin{tabular}[c]{@{}c@{}}Very\\low\end{tabular} & Low  & Mild & High & Avg.           \\ 
\hline\hline
1                     & \multirow{4}{*}{\begin{tabular}[c]{@{}c@{}}Manual\\DNN \end{tabular}} & \xmark                  & \xmark                         & 62.37    & 26.84 & 16.47 & 7.55 & 25.87           \\ 
\cline{1-1}\cline{3-9}
2                     &                                                                       & \multirow{3}{*}{\cmark} & input                         & 63.51    & 27.93 & 19.84 & 9.75 & 27.79           \\
3                     &                                                                       &                        & 7th hidden                    & 63.19    & 26.60 & 16.80 & 8.23 & 26.28           \\
4                     &                                                                       &                        & score A+AV                    & \textbf{61.31}     & \textbf{25.64} & \textbf{15.94} & 7.68  & \textbf{25.28}$^\dag$  \\ 
\hline\hline
5                     & \multirow{4}{*}{\begin{tabular}[c]{@{}c@{}}NAS\\DNN \end{tabular}}  & \xmark                  & \xmark                         & 61.42    & 26.72 & 16.25 & 7.65 & 25.63           \\ 
\cline{1-1}\cline{3-9}
6                     &                                                                       & \multirow{3}{*}{\cmark} & input                         & 63.40    & 26.97 & 20.05 & 9.19 & 27.37           \\
7                     &                                                                       &                        & 7th hidden                    & 61.81    & 27.00 & 15.88 & 8.40 & 25.97           \\
8                     &                                                                       &                        & score A+AV                    & \textbf{60.30}     & \textbf{26.23} & \textbf{15.39} & 7.96  & \textbf{25.21}$^\dag$  \\
\hline\hline
\end{tabular}}
\vspace{-0.3cm}
\end{table}

The performance of various AVSR systems based on either the manually designed DNN or NAS auto-configured DNN architecture and constructed using the above cross-domain generated visual features are shown in lines 2 to 4, and 6 to 8 in Table~\ref{tab:audio_visual}. In these systems, three forms of audio-visual modalities fusion were used including: a) input audio-visual feature concatenation (Sys. 2, 6 in Table~\ref{tab:audio_visual}, also shown in the left part of Fig.~\ref{fig:base_av}); b) hidden layer fusion performed by concatenating the visual features with outputs of the last non-bottleneck hidden layer (Sys. 3, 7 in Table~\ref{tab:audio_visual}, also shown in the right part of Fig.~\ref{fig:base_av}); and c) score fusion (Sys. 4, 8 in Table~\ref{tab:audio_visual}) via a linear interpolation over the output layer probability scores of the baseline audio-only ASR system (Sys. 1 or 5), and those of the hidden layer fusion based AVSR systems (Sys. 3 or 7) using equal weights. The results suggest that score fusion between ASR and AVSR systems (Sys. 4, 8) consistently produced statistically significant ($\alpha=0.05$) WER reductions of 0.4\%-0.6\% over the comparable audio-only ASR systems (Sys. 1, 5).

The lowest WER of 25.21\% was obtained using the NAS auto-configured DNN AVSR system (Sys. 8 in Table~\ref{tab:audio_visual}, again shown in the last line in Table~\ref{tab:compare_previous}). To the best of our knowledge, this is the lowest WER published so far on the UASpeech test set of 16 dysarthric speakers reported in the literature. Performance contrasts between this system against previously published systems on the same task are shown in Table~\ref{tab:compare_previous}. In particular, compared with our CUHK 2018 system featuring a 6-way DNN system combination~\cite{yu2018development} which defined state-of-the-art performance at the time, an overall WER reduction of 5.39\% absolute (17.61\% relative) was obtained. Furthermore, if excluding the 4 dysarthric speakers of very low intelligibility, the average WER obtained using our final AVSR system (Sys. 8, Table~\ref{tab:audio_visual}) is 15.79\%, close to the WERs found on normal speech recognition tasks.

\begin{table}[t]
\centering
\caption{Performance comparison between various recently published systems’ WERs on the UASpeech test set of 16 dysarthric speakers and our best system in this paper (Sys. 8, Table~\ref{tab:audio_visual})}
\vspace{-0.1cm}
\label{tab:compare_previous}
\scalebox{0.8}{\begin{tabular}{c|c} 
\hline\hline
Systems                                            & WER\%             \\ 
\hline
Sheffield-2013 Cross domain augmentation~\cite{christensen2013combining}            & 37.50           \\
CUHK-2021 LAS + CTC + Meta-learning + SAT \cite{wang2021improved}
                        & 35.00 \\
Sheffield-2015 Speaker adaptive training~\cite{sehgal2015model}           & 34.80           \\
Shefﬁeld-2020 Fine-tuning CNN-TDNN speaker adaptation~\cite{xiong2020source} & 30.76 \\
CUHK-2018 DNN system combination~\cite{yu2018development}                   & 30.60           \\
CUHK-2021 QuartzNet + CTC + Meta-learning + SAT~\cite{wang2021improved} & 30.50 \\
Sheffield-2019 Kaldi TDNN + Data Aug.~\cite{xiong2019phonetic}               & 27.88           \\
CUHK-2020 Cross-domain AVSR~\cite{liu2020exploiting} & 26.84 \\
CUHK-2020 DNN + Data Aug. + LHUC SAT~\cite{geng2020investigation}                     & 26.37           \\
\textbf{NAS DNN + Data Aug. + LHUC SAT + AV fusion (ours)} & \textbf{25.21}  \\
\hline\hline
\end{tabular}}
\vspace{-0.3cm}
\end{table}

\vspace{-0.1cm}
\section{Experiments on the Cantonese CUDYS corpus}

In this section, a comparable set of modelling components and techniques that previously featured in the best performing systems on the English UASpeech task: neural architecture search based DNN auto-configuration of Sec.~\ref{sec:asr_design}, speed perturbation based data augmentation of Sec.~\ref{sec:data_aug} and LHUC speaker adaptive training of Sec.~\ref{sec:spkr_adapt}, were further evaluated on a Cantonese CUDYS dysarthric speech corpus~\cite{wong2015development}.

The original 10-hour CUDYS corpus was further enlarged with more dysarthric speech collected since its initial release in 2015, and now contains speech from 27 impaired speakers. The development and evaluation sets, which were derived from a subset of 3.6 hours of speech collected from 21 impaired speakers and based on short sentences, were used for performance evaluation. The remaining part of the CUDYS data, after being further supplemented with normal Cantonese speech data from the SpeechOcean collection\footnote{http://en.speechocean.com/datacenter/recognition.html}, formed a baseline training data set of 21.4 hours. After speed perturbation based data augmentation techniques of Sec.~\ref{sec:data_aug} was applied, the training data size was further increased to 33.9 hours. In contrast to the UASpeech task based on single word utterances, each utterance in this task contains an average of more than six characters. Hence, a manually configured lattice-free MMI~\cite{povey2016purely} trained factorized TDNN (f-TDNN) baseline system~\cite{waibel1989consonant, peddinti2015time, povey2018semi, povey2016purely} with 7 context-splicing layers was used to model longer acoustic contexts. 40-dimension Mel-scale filter banks together with pitch parameters were used as the inputs features for system development. The Gumble-Softmax DARTS based neural architecture search approach of Sec.\ref{sec:asr_design} was then applied to automatically learn the left and right context offsets\footnote{Maximum context offset is set to be 6 for both left and right in each layer.} and the linear projection layer dimensionality\footnote{Searched over 6 different choices of projection dimensions \{100,120,160,200,240,300\}.} of each factored TDNN hidden layer. Speaker level variability was modelled using LHUC SAT and test time unsupervised adaptation. Due to the poor quality of video recordings in the CUDYS corpus caused by non-frontal face poses, and the difficulty in accessing high quality Cantonese audio-visual normal speech corpora with accurate transcripts required for the MLAN cross-domain visual feature generation approach of Sec.~\ref{sec:avsr}, experiments were conducted on audio-only ASR systems for the CUDYS task. A 4-gram language model with a 80K vocabulary was used. The character error rate (CER) metric was used for performance evaluation. 

\begin{table}[t]
\vspace{-0.3cm}
\centering
\caption{Description and performance of manually designed, NAS auto-configured LF-MMI trained factored TDNN systems and various end-to-end systems constructed on the CUDYS corpus. "CTL" means perturbing the healthy speech towards "disorder like" speech and "DYS" means perturbing the existing dysarthric speech data.}
\label{tab:CUDYS}
\scalebox{0.6}{\begin{tabular}{c|c|c|c|c|c|c|c|c|c|c|c} 
\hline\hline
\multirow{3}{*}{Sys.} & \multirow{3}{*}{Model}                                             & \multirow{3}{*}{Tgt.} & \multirow{3}{*}{\# Para} & \multirow{3}{*}{\begin{tabular}[c]{@{}c@{}}LHUC\\SAT\end{tabular}} & \multicolumn{2}{c|}{\multirow{2}{*}{Data Aug.}} & \multicolumn{5}{c}{CER\%}                                                     \\ 
\cline{8-12}
                      &                                                                    &                       &                          &                           & \multicolumn{2}{c|}{}                           & \multicolumn{2}{c|}{dev} & \multicolumn{2}{c|}{eval} & \multirow{2}{*}{Avg.}  \\ 
\cline{6-11}
                      &                                                                    &                       &                          &                           & CTL   & DYS                                     & Low  & High              & Low  & High               &                        \\ 
\hline
1                     & \multirow{4}{*}{TDNN}                                              & \multirow{4}{*}{phn.} & \multirow{3}{*}{9.8M}    & \multirow{3}{*}{\xmark}    & \multicolumn{2}{c|}{\xmark}                      & 88.0 & 13.4              & 73.6 & 1.4                & 19.6                   \\ 
\cline{6-12}
2                     &                                                                    &                       &                          &                           & \xmark & \cmark                                   & 85.5 & 7.1               & 66.1 & 1.1                & 15.8                   \\ 
\cline{6-12}
3                     &                                                                    &                       &                          &                           & \multicolumn{2}{c|}{\multirow{2}{*}{\cmark}}                      & 85.3 & 5.0               & 70.4 & 0.7                & 15.4                   \\ 
\cline{4-5}\cline{8-12}
4                     &                                                                    &                       & 10.1M                    & \cmark                     & \multicolumn{2}{c|}{}                      & 76.9 & 1.9               & 63.1 & 0.6                & 12.7                   \\ 
\hline
5                     & \multirow{2}{*}{\begin{tabular}[c]{@{}c@{}}NAS\\TDNN\end{tabular}} & \multirow{2}{*}{phn.} & 10.2M                    & \xmark                     & \multicolumn{2}{c|}{\multirow{2}{*}{\cmark}}     & 81.5 & 4.1               & 60.2 & 0.6                & 13.5                   \\
6                     &                                                                    &                       & 10.6M                    & \cmark                     & \multicolumn{2}{c|}{}                           & 71.9 & 3.2               & 50.2 & 0.5                & 11.2                   \\ 
\hline\hline
7                     & CTC                                                                & phn.                  & \multirow{4}{*}{9.8M}    & \multirow{4}{*}{\xmark}    & \multicolumn{2}{c|}{\multirow{4}{*}{\cmark}}     & 87.3 & 16.9              & 78.1 & 9.1                & 25.1                   \\
8                     & LAS                                                                & char.                 &                          &                           & \multicolumn{2}{c|}{}                           & 81.7 & 10.4              & 72.0 & 8.5                & 20.9                   \\
9                     & \begin{tabular}[c]{@{}c@{}}Pychain\\TDNN\end{tabular}              & phn.                  &                          &                           & \multicolumn{2}{c|}{}                           & 82.0 & 6.6               & 76.2 & 3.1                & 17.7                   \\
\hline\hline
\end{tabular}}
\vspace{-0.3cm}
\end{table}

The performance comparison between the baseline manually configured TDNN, NAS auto-configured DNN, before and after data augmentation and LHUC SAT were applied, and further against the comparable graphemic (character) LAS, phonetic CTC and Pychain TDNN systems are shown in Table~\ref{tab:CUDYS}. The same trends as previously observed on the English UASpeech task in Sec.~\ref{sec:asr_design} to Sec.~\ref{sec:spkr_adapt} can be found. First, data augmentation reduced the CER by a statistically significant ($\alpha=0.05$) margin of 4.2\% absolute when the baseline LF-MMI TDNN system was retrained using the larger augmented data set of 33.9 hours (Sys. 3 vs. Sys. 1). Second, NAS auto-configured TDNN\footnote{NAS selected projection dimensions at each layer: \{160,160,100,100,120,160,300\} and context configurations \{-4,6\},\{-5,4\},\{-6,6\},\{-6,6\},\{-6,6\},\{-6,6\},\{-6,6\}. $\eta$ in Eqn.~\ref{eq:penalized_darts} is set to be 0.} also reduced the CER significantly ($\alpha=0.05$) by 1.9\% absolute prior to speaker adaptation being applied (Sys. 5 vs. Sys. 3). The improvement from the NAS auto-configured TDNN system was retained after LHUC SAT and unsupervised speaker adaptation (Sys. 6 vs. Sys. 4). Large and statistically significant ($\alpha=0.05$) CER differences were found between the best performing NAS auto-configured TDNN system (Sys. 6) and the end-to-end systems of similar complexity.
\section{Conclusion}
\label{sec:conclusion}


This paper presented a series of developments associated with the design of state-of-the-art dysarthric speech recognition systems on the largest publicly available UASpeech dysarthric English speech corpus and a Cantonese CUDYS dysarthric speech dataset. Experimental results suggest the following trends. 

First, the suitability of current data-intensive deep learning based speech recognition system architectures, for example, end-to-end systems that traditionally benefit from the use of large quantities of data, needs to be re-assessed when being applied to dysarthric speech recognition tasks. This is due to the limited data quantity resulted from the difficulty in impaired speech data collection, and the 
large mismatch against normal speech. To this end, the auto-configured model structures derived from neural architecture search have been shown to produce better performance than a range of expert designed or manually configured systems of comparable or larger model complexity, before and after data augmentation or domain adaptation is used. Second, data augmentation techniques can effectively expand the limited training data by taking into account the systematic spectral and temporal deviation of dysarthric speech from normal speech. Third, the proposed speaker adaptation techniques can model the large variability among impaired speakers in both the original and augmented data, as well as allow fast adaptation to individual dysarthric speakers to be effectively performed using as little as a few seconds of speech. This user-centric feature is important when practically deploying speech recognition based assistive technologies to serve such people. Lastly, the use of visual features can further improve the recognition performance particularly for impaired speakers of low intelligibility whose voice quality is severely degraded. 

The combination of these techniques produced the lowest published word error rate (WER) of 25.21\% on the UASpeech test set 16 dysarthric speakers, and an overall WER reduction of 5.39\% absolute (17.61\% relative) over a very complex CUHK 2018 dysarthric speech recognition system using a 6-way DNN system combination and cross adaptation of out-of-domain normal speech data trained systems. Similar trends of performance improvements obtained using these techniques were also found on the CUDYS Cantonese dysarthric speech recognition task. The average WER over dysarthric speakers on the English UASpeech task obtained by our best AVSR system, if excluding the most difficult speakers of very low intelligibility, is 15.79\%. This is considered to be close to the WERs often found on normal speech recognition tasks. Future research will focus on designing neural network architectures and multi-modal speech recognition systems suitable for dysarthric speakers of very low intelligibility.

\section*{Acknowledgment}
We thank Disong Wang for sharing their cross-domain LAS system results. This research is supported by Hong Kong Research Grants Council GRF grant No. 14200218, 14200220, Theme based Research Scheme T45-407/19N, Innovation \& Technology Fund grant No. ITS/254/19, PiH/350/20, InP/275/20, and Shun Hing Institute of Advanced Engineering grant No. MMT-p1-19.

\ifCLASSOPTIONcaptionsoff
  \newpage
\fi

\bibliographystyle{IEEEtran}
\bibliography{IEEEabrv}

\end{document}